\documentclass[aps,twocolumn,superscriptaddress,longbibliography,floatfix,notitlepage]{revtex4-2}
\usepackage{bbm}
\usepackage{physics}
\usepackage{amsmath}
\usepackage{epsfig}
\usepackage{subfigure,mathrsfs}
\usepackage{array}
\usepackage{amssymb}
\usepackage{braket}
\usepackage{lmodern,amssymb}
\usepackage{physics}
\usepackage[dvipsnames]{xcolor}
\usepackage{lipsum}

\newcommand{\beq}[0]{\begin{equation}}
\newcommand{\eeq}[0]{\end{equation}}

\def\be{\begin{equation}}
\def\ee{\end{equation}}
\def\bea{\begin{eqnarray}}
\def\eea{\end{eqnarray}}
\newcommand{\ba}{\begin{eqnarray}}
\newcommand{\ea}{\end{eqnarray}}
\usepackage{hyperref}

\usepackage[bb=boondox]{mathalfa}

\begin{document}

\title{Inverse Current in Coupled Transport: A Quantum Thermodynamic Model}

\author{Shuvadip Ghosh}
\thanks{shuvadipg21@iitk.ac.in}
\affiliation{Indian Institute of Technology Kanpur, 
	Kanpur, Uttar Pradesh 208016, India}
\author{Nikhil Gupt}
\affiliation{Indian Institute of Technology Kanpur, 
	Kanpur, Uttar Pradesh 208016, India}
\affiliation{Department of Chemistry, University of Pennsylvania, Philadelphia, Pennsylvania 19104, USA}
\author{Arnab Ghosh}
\affiliation{Indian Institute of Technology Kanpur, 
	Kanpur, Uttar Pradesh 208016, India}

\begin{abstract}
The recent discovery of inverse current in coupled transport (ICC) in classical systems~\textcolor{blue}{[\textbf{Phys. Rev. Lett.} \textbf{124}, 110607 (2020)]} ---where an induced current flows opposite to two mutually parallel thermodynamic forces, yet remains consistent with the second law of thermodynamics --- reveals a striking and counterintuitive transport phenomenon. Using an exactly solvable model of strongly coupled quantum dots, we develop a thermodynamic framework to describe the ICC phenomenon at the quantum level. By systematically connecting the microscopic and macroscopic formulations of the entropy production rate in terms of appropriate entropic biases and entropic fluxes, our analysis identifies the conditions under which a \textit{genuine} ICC effect can arise in quantum thermal transport and highlights potential applications in autonomous quantum engines and refrigerators.
\end{abstract}

\maketitle

\section{Introduction}

\par Force and flux are two central thermodynamic quantities, interlinked through a cause-and-effect relationship. A thermodynamic force ($\mathcal{F}$) engenders a response in the form of a flux ($J$) within the system, causing a departure from equilibrium. For a single driving force, the resulting flux always aligns with the direction of the applied force, establishing a new equilibrium, consistent with the positive entropy production rate $\dot{\Sigma}=J\mathcal{F}\geq 0$~\cite{callen1985book,kondepudi2015book}. The situation changes significantly in the presence of multiple force-flux pairs. For example, a thermal (energy) force $\mathcal{F}_{\rm{E}}$ and a particle force $\mathcal{F}_{\rm{N}}$ can drive both particle $(J_{\rm{N}})$ and energy $(J_{\rm{E}})$ currents. Consequently, the rate of entropy production near equilibrium is expressed as $\dot{\Sigma}=J_{\rm{E}}\mathcal{F}_{\rm{E}}+J_{\rm{N}}\mathcal{F}_{\rm{N}}$, where $J_{\rm{E}}(J_{\rm{N}})$ is the thermodynamic flux conjugate to the thermodynamic force $\mathcal{F}_{\rm{E}}(\mathcal{F}_{\rm{N}})$. This could lead to the possibility of \textit{Inverse Current in Coupled transport} (ICC) when both forces are \textit{mutually parallel}, yet one of the induced currents flows \textit{against} both forces. Although the sign of a current is a matter of convention, for ICC to occur, a current must flow simultaneously against both \textit{mutually parallel} thermodynamic forces. Consequently, we set without loss any generality, both $\mathcal{F}_{\rm{E}}>0$ and $\mathcal{F}_{\rm{N}}>0$ in $\dot{\Sigma}=J_{\rm{E}}\mathcal{F}_{\rm{E}}+J_{\rm{N}}\mathcal{F}_{\rm{N}}$, so that a negative current signals ICC~\cite{wang2020inverse}. An immediate consequence is that simultaneous ICC in both fluxes is impossible, as it would violate the second law of thermodynamics. This is the result of Wang \textit{et. al.}~\cite{wang2020inverse} who recently demonstrated the novel effect of ICC in classical systems. Although the phenomenon appears highly counterintuitive, it is not forbidden, as long as the overall entropy production rate remains non-negative. Following the results of Wang \textit{et. al.}~\cite{wang2020inverse}, multiple attempts have been made to extend the concept of the ICC to quantum systems with coupled quantum dots (QDs)~\cite{zhang2021inverse,zhang2023inverse}. Although Refs.~\cite{zhang2021inverse,zhang2023inverse} numerically obtained the ICC effect, a rigorous analytical formulation in terms of well-defined thermodynamic forces and their corresponding fluxes is still missing for such coupled QD systems.

For this purpose, we propose an exactly solvable model of coupled QDs to demonstrate the ICC effect. Though our model is based on a three-terminal setup, we show that it can be effectively reduced to a \textit{two-terminal configuration}, essential for \textit{uniquely} identifying coupled transport driven by pairs of thermodynamic forces and their conjugate fluxes [see Fig.~\ref{Fig.3}]. However, the three-terminal structure remains indispensable because the QDs are capacitively coupled. Consequently, to generate both energy and particle currents within a minimal setup, one of the QDs must be simultaneously coupled to multiple thermal reservoirs. We further emphasize that interdot interactions play a crucial role in establishing ICC behavior under the near-equilibrium condition. Most importantly, we derive the conditions for genuine ICC in both energy and (spin-polarized) particle currents, fully consistent with the second law of thermodynamics.

The structure of the paper is as follows: Section~\ref{Sec-II} outlines our model based on coupled QDs and presents the system dynamics. In Sec.~\ref{Sec-IV}, we analyze steady-state currents and examine entropy production, including both macroscopic and microscopic expressions for the entropic forces and fluxes. Section~\ref{Sec-VI} summarizes the differences between the novel ICC phenomenon and the normal cross effects observed in conventional QD thermoelectric setups. In Sec.~\ref{Sec-VIII}, we present our results and discuss genuine ICC in energy and spin-polarized particle currents. Finally, we conclude in Sec.~\ref{Sec-IX}.

\section{Model and System Dynamics}\label{Sec-II}

\begin{figure}
    \centering    \includegraphics[width=\columnwidth,height=5.3cm]{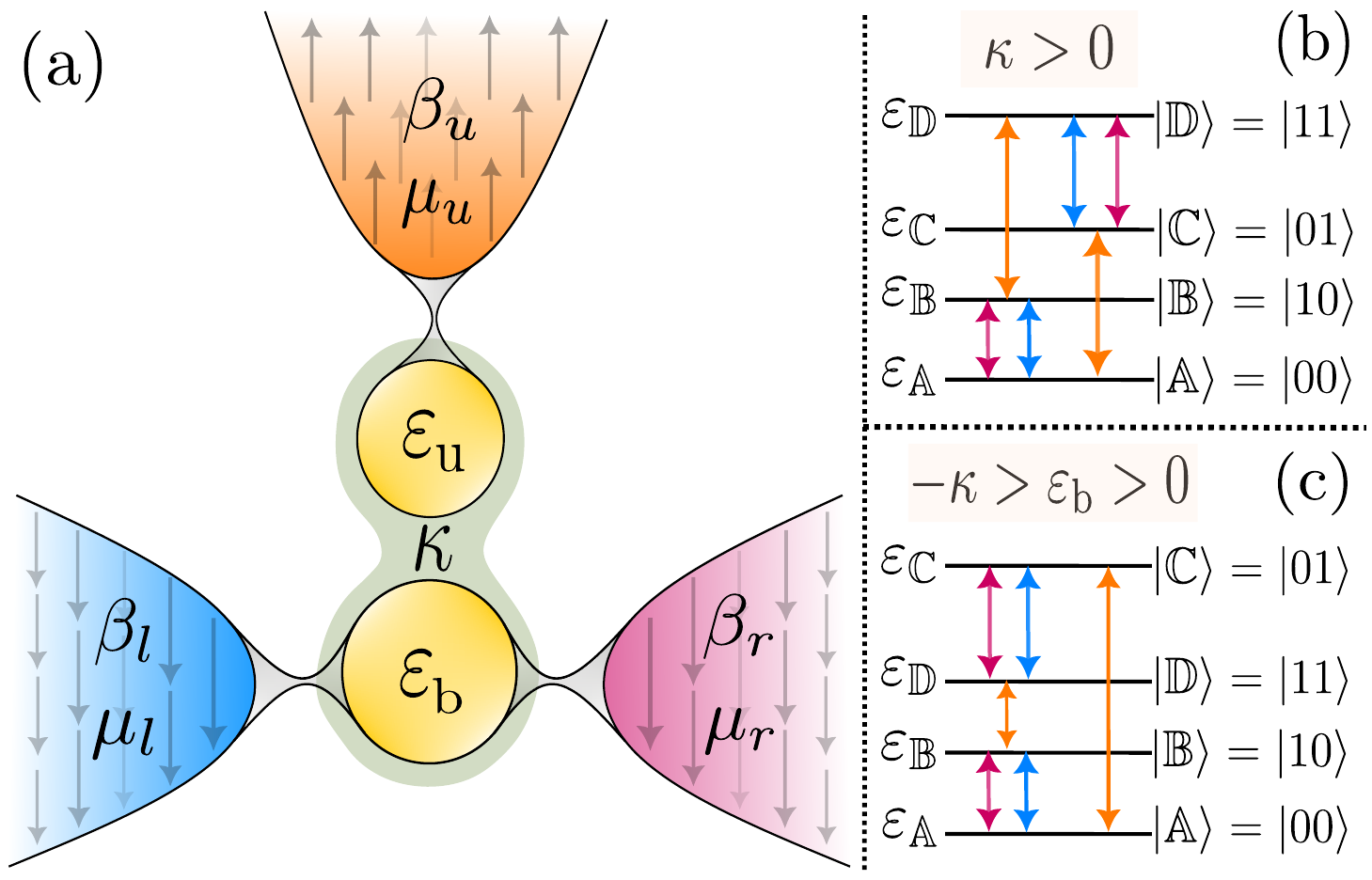}
    \caption{(a) Schematic diagram of the three-terminal Coulomb-coupled QDs. The energy level diagram of the eigenstates of coupled QDs for (b) $\kappa>0$ and (c) $\kappa<0;|\kappa|>\varepsilon_{\rm{b}}$ i.e. $-\kappa>\varepsilon_{\rm{b}}>0$. For specific values of bath parameters $\{\beta, \mu\}$, the above model reduces to a coupled transport phenomenon with a pair of thermodynamic forces [Fig.~\ref{Fig.3}].}
    \label{Model}
\end{figure}

Our model consists of two quantum dots (${\rm QDs}$) that are strongly and capacitively coupled, labeled upper (${\rm QD_u}$) and bottom (${\rm QD_b}$), as illustrated in [FIG.~\ref{Model}(a)]. The dots interact via a long-range Coulomb force~\cite{gupt2024graph,shuvadip2022univarsal,strasberg2022quantum}, preventing particle exchange while allowing energy exchange due to Coulomb interaction ($\kappa_{\rm{c}}$). Since inter-dot particle hopping is restricted, generating particle current requires at least one QD to be simultaneously coupled to more than one reservoir~\cite{harbola2006quantum,sanchez2011optimal,strasberg2022quantum}.  We assume that ${\rm QD_b}$ is tunnel-coupled to two fermionic leads, labeled as $l$ (\textit{left}) and $r$ (\textit{right}), respectively. The other dot, ${\rm QD_u}$, is tunnel-coupled to only a single reservoir, labeled as $u$. For spinless electrons, inter-dot interaction is always positive; however, we extend our analysis to spin-polarized~\cite{gupt2024graph,wang2022cycleflux} electrons, where, we assume that ${\rm QD_b}$ is coupled to spin-down ($\downarrow$) fermionic reservoirs, and ${\rm QD_u}$ to spin-up ($\uparrow$) one [FIG.~\ref{Model}(a)]. The Hamiltonian of the coupled ${\rm QD}$ system is given by
\begin{equation}\label{HS}
\begin{split}
 H_{\rm{S}}=\varepsilon_{\rm{b}} \mathcal{N}_{\rm{b}\downarrow}+\varepsilon_{\rm{u}} \mathcal{N}_{\rm{u}\uparrow}+\kappa_{c}\mathcal{N}_{\rm{b}\downarrow}\mathcal{N}_{\rm{u}\uparrow} + \kappa_{s}\sigma^{z}_{\rm{b}\downarrow}\sigma^{z}_{\rm{u}\uparrow}.
\end{split}
\end{equation}
In Eq.~\eqref{HS}, $\varepsilon_\alpha>0$ $(\alpha= \rm{u},\rm {b})$ denotes the single-particle energy level of the $\alpha$'th ${\rm QD}$, which are taken to be positive without loss of generality. Since the electron density in the dots is low, the occupancy limits to either zero or one~\cite{whitney2018quantum,gupt2023topranked,gupt2024graph,shuvadip2022univarsal}. Thus eigenstates of ${\rm QD_{\rm{b}(\rm{u})}}$ are $|0\rangle$ and $\ket\downarrow(\ket\uparrow)$, with energy eigenvalues $0$ and $\varepsilon_{\rm{b}} (\varepsilon_{\rm{u}})$, respectively. The number operators are $\mathcal{N}_{\rm{b}\downarrow}= d^\dagger_{\rm{b}\downarrow} d_{\rm{b}\downarrow}$ and $\mathcal{N}_{\rm{u}\uparrow}=d^\dagger_{\rm{u}\uparrow} d_{\rm{u}\uparrow}$, where $d^\dagger_{\alpha\sigma}$ ($d_{\alpha\sigma}$) represent the electron creation and annihilation operators with spin $\sigma = \{\uparrow, \downarrow \}$, obeying $\{d_{\alpha\sigma},d^\dagger_{\alpha^{\prime}\sigma^{\prime}}\}=\delta_{\alpha \alpha^{\prime}}\delta_{\sigma \sigma^{\prime}}$. The spin-spin interaction term~\cite{daroca2025role} in Eq.~\eqref{HS} is expressed as $\sigma^{z}_{\rm{b\downarrow}}=1- 2\mathcal{N}_{\rm{b\downarrow}}$ and $\sigma^{z}_{\rm{u\uparrow}}=2\mathcal{N}_{\rm{u\uparrow}}-1$, where the spin operator $\sigma^{z}$ satisfies the relations $\sigma^{z}_{\rm{b}\downarrow}|\downarrow \rangle = -1 |\downarrow \rangle$ and $\sigma^{z}_{\rm{u}\uparrow}|\uparrow \rangle = +1 |\uparrow \rangle$.

With strong capacitive coupling~\cite{wang2022cycleflux,gupt2024graph,shuvadip2022univarsal}, $H_{\rm{S}}$ is diagonal in the tensor product eigenbasis of the individual ${\rm QD}$ number operators. Thus four eigenstates $\{|0\rangle, \ket{\downarrow}\}\otimes\{|0\rangle, \ket{\uparrow}\}$,  are labeled by $|\mathbb{A}\rangle=\ket{00}$, $|\mathbb{B}\rangle=\ket{\downarrow 0}$, $|\mathbb{C}\rangle=\ket{0\uparrow}$, $|\mathbb{D}\rangle=\ket{\downarrow\uparrow}$ with corresponding energies ($\varepsilon_\mathbb{i}$, $\mathbb{i=A,B,C,D}$):  $\varepsilon_\mathbb{A}=0$, $\varepsilon_\mathbb{B}=\varepsilon_{\rm{b}}$, $\varepsilon_\mathbb{C}=\varepsilon_{\rm{u}}$ and $\varepsilon_\mathbb{D}=\varepsilon_{\rm{b}}+\varepsilon_{\rm{u}}+\kappa$ [FIG.~\ref{Model}(b),~\ref{Model}(c)]. We assume, w.l.o.g, that $\varepsilon_{\rm{b}} < \varepsilon_{\rm{u}}$. The total interaction, $\kappa=\kappa_{\rm{c}}-\kappa_{\rm{s}}$, can be either attractive ($\kappa<0$) or repulsive ($\kappa>0$), while $\kappa_{\rm{c}}$ and $\kappa_{\rm{s}}$ are both positive. If $\kappa<0$ and $|\kappa|>\varepsilon_{\rm b}$, the energy states $|\mathbb{C}\rangle$ and $|\mathbb{D}\rangle$ interchange their positions~[FIG.~\ref{Model}(c)]. Although an isolated electron produces a repulsive Coulomb potential, if it is embedded in a medium that flips the sign of its Coulomb potential, it could become attractive to other electrons~\cite{hamo2016electron,tabatabaei2018charge,little1964possibility,prawiroatmodjo2017negativeU}. Hamo et al.~\cite{hamo2016electron} recently demonstrated such attractions in a carbon-nanotube double QD coupled to a charge qubit. In their setup, the double QD and the qubit are fabricated on separate microchips and aligned perpendicularly. When the qubit is brought sufficiently close, it induces attraction between electrons in the double QD. The oscillating polarization field of the qubit ``dresses" the electron potentials, favoring doubly occupied states over singly occupied ones~\cite{hamo2016electron,tabatabaei2018charge,little1964possibility,prawiroatmodjo2017negativeU}, as shown in FIG.~\ref{Model}(c). Owing to its high degree of tunability, the above platform is well suited for realizing ICC in coupled QDs, as discussed in Sec.~\ref{Sec-VIII}.

Finally, electron (Fermionic) reservoirs are characterized by temperature and chemical potential. The Hamiltonian for the $\lambda$'th spin-polarized bath with spin $\sigma = \{\uparrow, \downarrow \}$, is defined as $H_{\rm{B}}^{\lambda}\equiv H_{\rm{B}}^{\lambda\sigma}=\sum_{k} (\epsilon^{\lambda\sigma}_{k}-\mu_{\lambda\sigma})c_{{\lambda\sigma} k}^\dagger c_{{\lambda\sigma} k}$, where $\epsilon^{\lambda\sigma}_k$ is the energy of the non-interacting electrons, $\mu_{\lambda\sigma}$ is the chemical potential, and $c^\dagger(c)$ are creation (annihilation) operators, obeying anti-commutation relation~\cite{nikhil2021statistical,samarth2023introduction,ghosh2012fermionic,sinha2011decay,sinha2011quantum,damas2023cooling}. Finally, QDs are weakly coupled to the reservoirs, allowing sequential tunneling ~\cite{shuvadip2022univarsal,gupt2023topranked,gupt2024graph,dutta2017thermal,dutta2020single,thierschmann2015three}, where one ${\rm QD}$ interacts with one lead at a time. The tunnel-coupled Hamiltonians characterized by the coupling constants $t_{k}^{\alpha\sigma\lambda}$ are given by 
\begin{eqnarray}\label{HT}
 H_{\rm{T}}^{{\rm{b}\downarrow}{l(r)}}&=&\hbar\sum_k[t^{{\rm{b}\downarrow}{l(r)}}_k c_{l(r)\downarrow k}^\dagger d_{{\rm{b}}\downarrow}+t^{{\rm{b}\downarrow}{l(r)*}}_ kd_{{\rm{b}}\downarrow}^{\dagger}c_{l(r)\downarrow k}],\nonumber\\
H_{{\rm{T}}}^{{\rm{u}\uparrow}{u}}&=&\hbar\sum_k[t^{{\rm{u}\uparrow}{r}}_k c_{u\uparrow k}^\dagger d_{{\rm{u}}\uparrow}+t^{{\rm{u}\uparrow}{u*}}_ kd_{{\rm{u}}\uparrow}^{\dagger}c_{u\uparrow k}],   
\end{eqnarray}
where, $H_{\rm{T}}^{{\rm{b}\downarrow}{l(r)}}$ and $H_{\rm{T}}^{{\rm{u}\uparrow}{u}}$ describe the interaction between $\rm{QD_b}$ and $\rm{QD_u}$ with reservoirs $l(r)$ via $\downarrow$ electrons and reservoir $u$ via $\uparrow$ electrons, respectively. Coulomb blockade and the sequential tunneling approximation prohibit transitions $|\mathbb{B}\rangle\leftrightarrow|\mathbb{C}\rangle$ and $|\mathbb{A}\rangle\leftrightarrow|\mathbb{D}\rangle$, allowing four allowed transitions. Reservoirs $l$, $r$ control $|\mathbb{A}\rangle\leftrightarrow|\mathbb{B}\rangle$ and $|\mathbb{C}\rangle\leftrightarrow|\mathbb{D}\rangle$, while $u$ governs $|\mathbb{A}\rangle\leftrightarrow|\mathbb{C}\rangle$ and $|\mathbb{B}\rangle\leftrightarrow|\mathbb{D}\rangle$ as designated in FIG.~\ref{Model}(b),~\ref{Model}(c). The transition energies are defined as $\omega_\mathbb{ij}=\varepsilon_\mathbb{j}-\varepsilon_\mathbb{i}$, with $\omega_{\mathbb{AB}}=\varepsilon_{\rm{b}}$, $\omega_{\mathbb{AC}}=\varepsilon_{\rm{u}}$, $\omega_{\mathbb{CD}}=\varepsilon_{\rm{b}}+\kappa$, and $\omega_{\mathbb{BD}}=\varepsilon_{\rm{u}}+\kappa$, [FIG.~\ref{Model}(b),~\ref{Model}(c)].

\par The reduced state of the composite system is represented by the density matrix  $\rho_{\rm{s}}(t)=\Tr_{\rm{B}}\{\rho_{\rm{tot}}(t)\}$, where $\rho_{\rm{tot}}(t)$ is the total density matrix of the system and bath combined. Using strong-coupling formalism, the time evolution of $\rho_{\rm{s}}(t)$ is governed by the Lindblad Master Equation (LME) under the Born, Markov, and Secular (BMS) approximations~\cite{breuer2002book,strasberg2022quantum} (Appendix-~\ref{Appendix-A}):
\begin{equation}\label{LME}
\frac{d}{dt}\rho_{\rm{s}}(t)=\sum_{\lambda}\mathcal{L}_{\lambda}[\rho_{\rm{s}}(t)]; \quad \lambda=l,r,u.
\end{equation}
It is worth noting that the strong coupling formalism pertains to the interaction between QDs, while we still assume the weak coupling between the system and its environment. This validates the use of the BMS approximation to derive Eq.~\eqref{LME} based on the eigenstates of $H_{\rm S}$ involving both QDs. Consequently, the dissipation of each QD depends on not only its bath coupling(s) but also on the inter-dot interactions, which are crucial for accurately modeling heat flow across the entire system~\cite{levy2014local}. The Lindblad super-operator $\mathcal{L}_\lambda[\rho_{\rm{s}}(t)]$ in Eq.~\eqref{LME} is defined as:
\begin{widetext}
\begin{equation}\label{Lindbladian}
\begin{split}
\mathcal{L}_{\lambda}[\rho_{\rm{s}}(t)]=& \sum_{\{\omega_{\alpha}\}>0}\left\{\gamma_\lambda(\omega_\alpha)f^+_\lambda(\omega_\alpha)\left[d_{\alpha\sigma}^\dagger(\omega_{\alpha})\rho_{\rm{s}} d_{\alpha\sigma}(\omega_{\alpha})-\frac{1}{2}{\{\rho_{\rm{s}}}, d_{\alpha\sigma}(\omega_{\alpha})d_{\alpha\sigma}^\dagger(\omega_{\alpha})\}\right] 
\right.\\&\left.+\gamma_\lambda(\omega_\alpha) {f}^-_\lambda(\omega_\alpha)
\left[d_{\alpha\sigma}(\omega_{\alpha})\rho_{\rm{s}} d_{\alpha\sigma}^\dagger(\omega_{\alpha})
-\frac{1}{2}{\{\rho_{\rm{s}}}, d_{\alpha\sigma}^\dagger(\omega_{\alpha})d_{\alpha\sigma}(\omega_{\alpha})\}\right]\right\}.
\end{split}
\end{equation}
\end{widetext}

\begin{figure}
\centering    \includegraphics[width=\columnwidth, height=4.3cm]{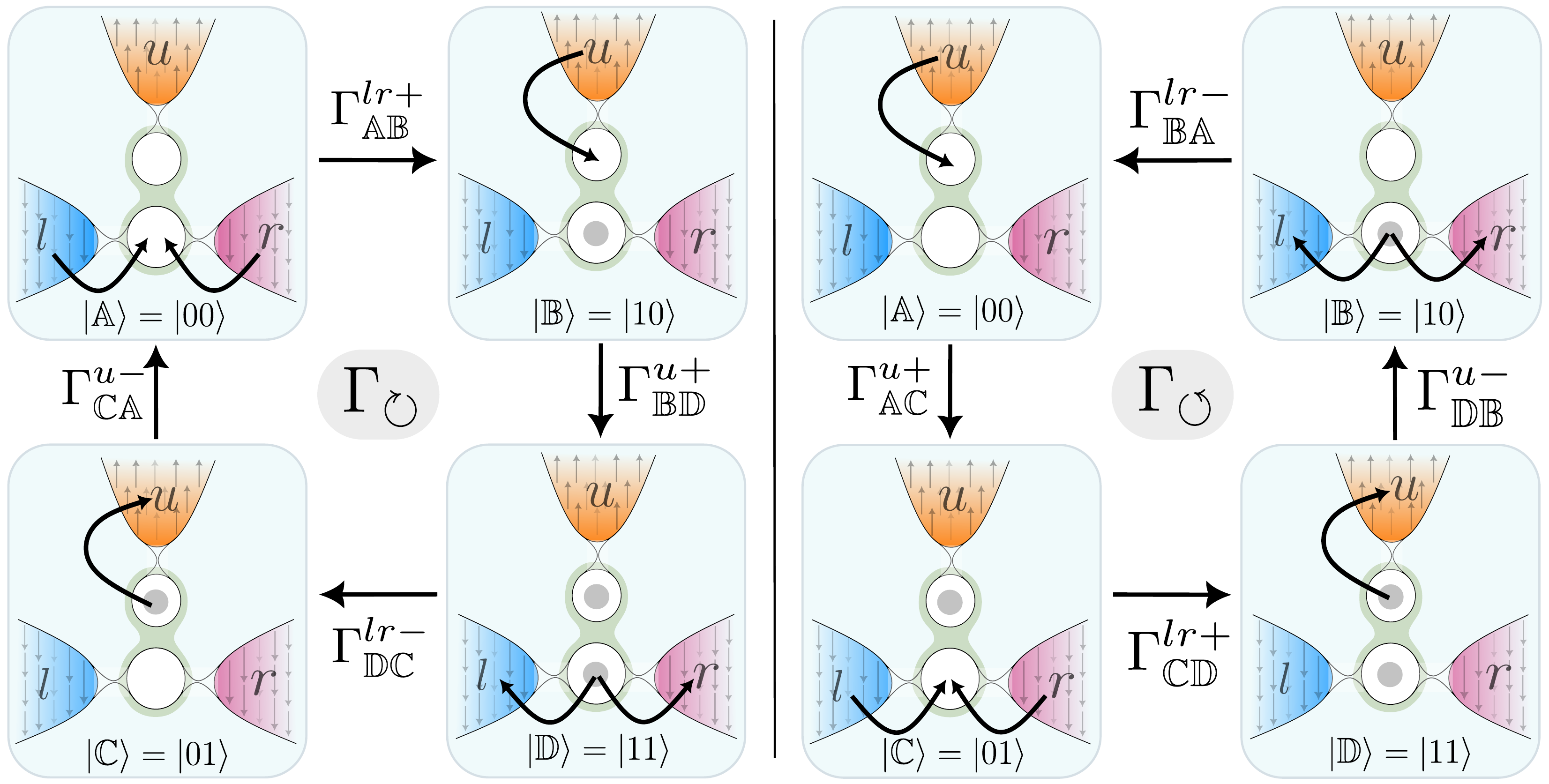}
\caption{Schematic representation of the clockwise (Left) and anti-clockwise (Right) transition cycles, induced by energy flow from one bath to another, associated with the particle exchange between QDs and the coupled reservoirs. For the Left (clockwise) cycle, the transition starts from the state $|\mathbb{A}\rangle = |00\rangle$ (where both QDs are vacant) to the state $|\mathbb{B}\rangle = |10\rangle$ (where the bottom QD is occupied), with a net transition rate $\Gamma_{\mathbb{AB}}^{lr+}$ (indicated above the arrow), induced by both the $l$ and $r$ reservoirs. Next, the system transitions from $|\mathbb{B}\rangle = |10\rangle$ to $|\mathbb{D}\rangle = |11\rangle$, where a spin-up electron enters the upper QD from reservoir $u$ with a net transition rate $\Gamma_{\mathbb{BD}}^{u+}$. The electron then leaves the bottom QD to either the $l$ or $r$ reservoir with rate $\Gamma_{\mathbb{DC}}^{lr-}$, changing the state 
from $|\mathbb{D}\rangle = |11\rangle$ to $|\mathbb{C}\rangle = |01\rangle$. Finally, the cycle 
closes when the electron in the upper QD leaves with rate $\Gamma_{\mathbb{CA}}^{u-}$, changing the state from $|\mathbb{C}\rangle = |01\rangle$ back to $|\mathbb{A}\rangle = |00\rangle$. A similar sequence occurs for the Right transition cycle, but in the reverse direction. Finally, at steady state, all transition rates of the clockwise (anti-clockwise) cycles are equal to each other and represented by $\Gamma_{\circlearrowright}(\Gamma_{\circlearrowleft})$ [Cf.~Eq.~\eqref{gamma}].}
        \label{Transition Cycle}
\end{figure} 
 
The bare electron transfer rate $\gamma_\lambda(\omega_\alpha)$ between reservoir $\lambda$ and ${\rm QD_\alpha}$ is given by Fermi's golden rule
$\gamma_\lambda(\omega_\alpha)\equiv\gamma_{\lambda\sigma}(\omega_\alpha)=2\pi \sum_{k} |t_{k}^{\alpha\sigma\lambda}|^2 \delta\big(\omega_\alpha-\epsilon_{k}^{\lambda\sigma}\big)$, where 
$\omega_\alpha$ is the energy associated with transitions between ${\rm QD_\alpha}$ and its coupled lead. For simplicity, we write $\gamma_{l(r)}(\omega_{\rm{b}})\equiv\gamma_{l(r)}$ and $\gamma_{u}(\omega_{\rm{u}})\equiv\gamma_{u}$, where $\omega_{\rm{b}}=\{\omega_{\mathbb{AB}},\omega_{\mathbb{CD}}\}$~and~$\omega_{\rm{u}}=\{\omega_{\mathbb{AC}},\omega_{\mathbb{BD}}\}$, for excitation (particle is entering into the system), and similarly for de-excitation (particle leaving from the system) [FIG.~\ref{Transition Cycle}]~\cite{sanchez2011optimal}. To simplify calculations without relying on the specific ordering of system eigen-energies, we define the Fermi distribution function (FDF), $f^{\pm}_\lambda(\omega_\mathbb{ij})$ governing transitions $|\mathbb{i}\rangle \rightarrow |\mathbb{j}\rangle$. The `$+$' sign represents excitation (particle entering the system), while `$-$' indicates de-excitation (particle leaving the system)~\cite{sanchez2011optimal}, which is explicitly evident from the FIG.~\ref{Transition Cycle}. This notation is independent of eigenstate arrangements and satisfies the relation~\cite{tesser2022heat}:
\begin{equation}\label{FDF}
f^{+}_\lambda(\omega_\mathbb{ij})=1-f^{-}_\lambda(\omega_\mathbb{ji})=\bigg[{1+\exp \bigg({\frac{\omega_\mathbb{ij}-\mu_\lambda}{k_BT_\lambda}}}\bigg)\bigg]^{-1},
\end{equation}
where $T_\lambda$ and $\mu_\lambda$ are the temperature and chemical potential of the $\lambda$-th  reservoir. From FIG.~\ref{Transition Cycle} and Eq.~\eqref{Lindbladian}, the rate equations for occupation probabilities  are:
\begin{equation}\label{rho2}
    \begin{split}
\dot{\rho}_\mathbb{A}=&-\Gamma^{lr+}_\mathbb{AB}-\Gamma^{u+}_\mathbb{AC}\quad;\quad  
\dot{\rho}_\mathbb{B}=\Gamma^{lr+}_\mathbb{AB}-\Gamma^{u+}_\mathbb{BD};
\\\dot{\rho}_\mathbb{C}=&-\Gamma^{lr+}_\mathbb{CD}+\Gamma^{u+}_\mathbb{AC}\quad;\quad\dot{\rho}_\mathbb{D}=\Gamma^{lr+}_\mathbb{CD}+\Gamma^{u+}_\mathbb{BD},
\end{split}
\end{equation}
where, $\dot{\rho}_\mathbb{i}=\sum_{\lambda=l,r,u}\langle \mathbb{i}|\mathcal{L}_{\lambda}[\rho_{\rm{s}}(t)]|\mathbb{i}\rangle$ and $\Gamma^{lr\pm}_\mathbb{ij}\equiv\Gamma^{l\pm}_\mathbb{ij}+\Gamma^{r\pm}_\mathbb{ij}$. The net transition rate $\Gamma^{\lambda\pm}_\mathbb{ij}$ mediated by reservoir $\lambda$ is expressed as:
\begin{equation}\begin{split}\label{ME14}
\Gamma_\mathbb{ij}^{\lambda\pm}\equiv\Gamma_\mathbb{i\mapsto j}^{\lambda\pm}
={\rm{k}}_\mathbb{ij}^{\lambda\pm}\rho_\mathbb{i}-{\rm{k}}_\mathbb{ji}^{\lambda\mp}\rho_\mathbb{j},
\end{split}
\end{equation}
where 
\begin{equation}\label{ME15}
{{\rm{k}}}_\mathbb{ij}^{\lambda\pm}=\gamma_{\lambda}f^{\pm}_{\lambda}(\omega_\mathbb{ij})\quad ;\quad  {{\rm{k}}}_\mathbb{ji}^{\lambda\mp}=\gamma_{\lambda}f^{\mp}_{\lambda}(\omega_\mathbb{ji}).
\end{equation}
For fermionic reservoirs: %${{\rm{k}}}_\mathbb{ij}^{\lambda\pm}+{{\rm{k}}}_\mathbb{ji}^{\lambda\mp}=\gamma_{\lambda}$.
\begin{equation}\label{k}
{{\rm{k}}}_\mathbb{ij}^{\lambda\pm}+{{\rm{k}}}_\mathbb{ji}^{\lambda\mp}=\gamma_{\lambda}.
\end{equation}
Assuming all $\gamma_\lambda$ are equal, we derive closed-form expressions for steady-state energy and particle currents in the following Section. As a final remark, we emphasize that the quantum master equation~\eqref{LME}-\eqref{Lindbladian} reduces to a rate equation [Eq.~\eqref{rho2}] that appears formally ``\textit{classical}'', yet retains intrinsic quantum-mechanical content. This is evident from the expressions for the net transition rates $\Gamma_{\mathbb{ij}}^{\lambda \pm}$, which depend on the Fermi--Dirac statistics of the associated electron reservoirs, the discrete energy levels of the coupled QDs, and the nature of the interaction between the spin-polarized electrons, which may be either attractive or repulsive. This has deeper implications, which is further illustrated at the end of Sec.~\ref{Sec-IV}.

\section{Steady State Currents: Entropic Representation}\label{Sec-IV}
\par To derive the expressions for steady-state currents under the grand canonical formalism, we consider the equilibrium initial density operator of the system $\rho_{\rm{s}}(0)$ as~\cite{strasberg2022quantum}
\begin{equation}   \rho_{\rm{s}}^{\rm{eq}}=\rho_{\rm{s}}(0)=\frac{e^{-\bar{\beta}(H_{\rm{s}}-\bar{\mu}\mathcal{N})}}{\mathcal{Z(\bar{\beta},\bar{\mu})}},
\end{equation}
where $\mathcal{Z(\bar{\beta},\bar{\mu})}=\Tr [{e^{-\bar{\beta}(H_{\rm{s}}-\bar{\mu}\mathcal{N})}}]$ is the grand canonical partition function and $\mathcal{N}=\mathcal{N}_{\rm{b}\downarrow}+\mathcal{N}_{\rm{u}\uparrow}$ being the total particle number operator of the two QDs. The $\bar{\beta}$ and $\bar{\mu}$ are the effective inverse temperature and chemical potential, $\bar{\beta}=\sum_{\lambda}\beta_{\lambda}$ and $\bar{\mu}=\sum_{\lambda}\mu_{\lambda}$ respectively. We assume that environmental interaction slightly perturbs the system from its initial equilibrium state $\rho_{\rm{s}}(0)=\rho_{\rm{s}}^{\rm{eq}}$ to $\rho_{\rm{s}}(t)$, such that $\delta\rho_{\rm{s}}(t)=\rho_{\rm{s}}(t)-\rho_{\rm{s}}(0)\equiv \mathcal{O}
(\xi)$, where $\xi$ is a small expansion parameter. Following Ref.~\cite{strasberg2022quantum}, equating system von-Neumann entropy (times $k_B$) with thermodynamic entropy near close to equilibrium~\cite{esposito2010entropy,landi2021irreversible} %$ \mathcal{S}_{\rm{s}}(t)=-k_B \Tr_{\rm{s}}[\rho_{\rm{s}}(t)\ln{\rho_{\rm{s}}(t)}]$
\begin{equation}\label{S1}
\begin{split}
\mathcal{S}_{\rm{s}}(t)=-k_B \Tr_{\rm{s}}[\rho_{\rm{s}}(t)\ln{\rho_{\rm{s}}(t)}],\\
\end{split}   
\end{equation}
and retaining only terms that are first order in $\xi$, the expression for $\varDelta\mathcal{S}_{\rm{s}}=\mathcal{S}_{\rm{s}}(t)-\mathcal{S}_{\rm{s}}(0)$ can be derived as~\cite{strasberg2022quantum}
\begin{equation}\label{S3}
\begin{split}
\varDelta\mathcal{S}_{\rm{s}}(t)
\approx&{-k_B \Tr_{\rm{s}}[\delta\rho_{\rm{s}}(t)\ln{\rho_{\rm{s}}^{\rm eq}}]}\\
=&k_B\beta\Tr_{\rm{s}}[\delta\rho_{\rm{s}}(t)H_{\rm{s}}]-k_B\beta\mu\Tr_{\rm{s}}[\delta\rho_{\rm{s}}(t)\mathcal{N}].
\end{split}
\end{equation} 
Comparing Eq.~\eqref{S3} and the first law of thermodynamics in the presence of \textit{chemical work done}, we identify change in energy as $\varDelta E=\Tr_{\rm{s}}[\delta\rho_{\rm{s}}(t)H_{\rm{s}}]$  and change in particles number as $\varDelta N=\Tr_{\rm{s}}[\delta\rho_{\rm{s}}(t)\mathcal{N}].$ 
Noting $\delta\rho_{\rm{s}}(t)=\rho_{\rm{s}}(t)-\rho_{\rm{s}}(0)$, and $H_{\rm{s}}$, $\mathcal{N}$ are time-independent, the net energy flux ($J_{\rm{E}}$) and particle flux ($J_{\rm{N}}$) are recognized as follows:
\begin{equation}\label{S6}
\begin{split}
J_{\rm{E}}(t)=&\Tr_{\rm{s}}[\Dot{\rho}_{\rm{s}}(t)H_{\rm{s}}]=\sum_{\lambda=a,b,r}\Tr_{\rm{s}}[\mathcal{L}_{\lambda}[\rho_{\rm{s}}(t)]H_{\rm{s}}],\\
J_{\rm{N}}(t)=&\Tr_{\rm{s}}[\Dot{\rho}_{\rm{s}}(t)\mathcal{N}]=\sum_{\lambda=a,b,r}\Tr_{\rm{s}}[\mathcal{L}_{\lambda}[\rho_{\rm{s}}(t)]\mathcal{N}],
\end{split}
\end{equation}
where we use Eq.~\eqref{LME} for $\Dot{\rho}_{\rm{s}}(t)$. Identifying $J_{\rm{E(N)}}$ as the sum of the contributions of energy (particle) flux  associated with all three reservoirs: $J_{\rm{E}}(t)=\sum_{\lambda}J^{\lambda}_{\rm{E}}(t)$; $J_{\rm{N}}(t)=\sum_{\lambda}J^{\lambda}_{\rm{N}}(t)$,
where $J^{\lambda}_{\rm{E(N)}}$ is positive if energy (particle) current flows from the reservoir $\lambda$ to the system, we obtain
\begin{equation}\label{S9}
J^{\lambda}_{\rm{E}}(t)=\Tr_{\rm{s}}[\mathcal{L}_{\lambda}[\rho(t)]H_{\rm{s}}] \;; \; 
J^{\lambda}_{\rm{N}}(t)=\Tr_{\rm{s}}[\mathcal{L}_{\lambda}[\rho(t)]\mathcal{N}].
\end{equation}

The heat current is obtained as $J^{\lambda}_{\rm{Q}}(t)=J^{\lambda}_{\rm{E}}(t)-\mu_{\lambda}J^{\lambda}_{\rm{N}}(t)$, which at steady state simplifies to $J^{\lambda}_{\rm{Q}}=J^{\lambda}_{\rm{E}}-\mu_{\lambda}J^{\lambda}_{\rm{N}}$. Here $J^{\lambda}_{\rm{E}}=\Tr_{\rm{s}}[\mathcal{L}_{\lambda}[\rho_{ss}]H_{\rm{s}}]$ and $J^{\lambda}_{\rm{N}}=\Tr_{\rm{s}}[\mathcal{L}_{\lambda}[\rho_{ss}]\mathcal{N}]$. Using Eq.~\eqref{Lindbladian} for the $\mathcal{L}_{\lambda}[\rho_{ss}]$, all three steady state currents are given by~\cite{gelbwaser2015thermodynamics,kurizki_kofman_2022}:
\begin{equation}\label{S15}
\begin{split}
J^{\lambda}_{\rm{E}}=&\sum_{\{\omega_{\mathbb{ij}}\}}{\omega_{\mathbb{ij}}}\Gamma_{\mathbb{ij}}^{\lambda+}=\sum_{\{\omega_{\mathbb{ji}}\}}{\omega_{\mathbb{ji}}}\Gamma_{\mathbb{ji}}^{\lambda-};\\
J^{\lambda}_{\rm{N}}=&\sum_{\{\omega_{\mathbb{ij}}\}}\Gamma_{\mathbb{ij}}^{\lambda+}=\sum_{\{\omega_{\mathbb{ji}}\}}\Gamma_{\mathbb{ji}}^{\lambda-};\\
J^{\lambda}_{\rm{Q}}=&\sum_{\{\omega_{\mathbb{ij}}\}}({\omega_{\mathbb{ij}}}-\mu_{\lambda})\Gamma_{\mathbb{ij}}^{\lambda+}=\sum_{\{\omega_{\mathbb{ji}}\}}({\omega_{\mathbb{ji}}}-\mu_{\lambda})\Gamma_{\mathbb{ji}}^{\lambda-}, 
\end{split}
\end{equation}
where we use the relation $\Gamma_{\mathbb{ij}}^{\lambda+}=-\Gamma_{\mathbb{ji}}^{\lambda-}$. The explicit expressions for all three currents are given in Appendix-~\ref{Appendix-B}. At steady state ($\dot{\rho}_\mathbb{i}=0$), the transition rates satisfy:
\begin{equation}\label{gamma}
\begin{split}
\Gamma^{lr+}_\mathbb{AB}=\Gamma^{u+}_\mathbb{BD}=\Gamma^{lr-}_\mathbb{DC}=\Gamma^{u-}_\mathbb{CA}=\Gamma_{\mathbb{ABCDA}}\equiv\Gamma_{\circlearrowright};\\
\Gamma^{u+}_\mathbb{AC}=\Gamma^{lr+}_\mathbb{CD}=\Gamma^{u-}_\mathbb{DB}=\Gamma^{lr-}_\mathbb{BA}=\Gamma_{\mathbb{ACDBA}}\equiv\Gamma_{\circlearrowleft},
\end{split}
\end{equation}
implying $\Gamma_{\circlearrowright}=-\Gamma_{\circlearrowleft}$ [FIG.~\ref{Transition Cycle}]. With the help of Eq.~\eqref{S15}, one can verify
\begin{equation}\label{Jer-Jnr}
\begin{split}
J_{\rm{E}}^u=&\varepsilon_{\rm{u}}\Gamma_{\circlearrowleft}+(\varepsilon_{\rm{u}}+\kappa)\Gamma_{\circlearrowright}=\kappa\Gamma_{\circlearrowright};
\\J_{\rm{N}}^u=&\Gamma_\mathbb{AC}^{u+}+\Gamma_\mathbb{BD}^{u+}=\Gamma_{\circlearrowleft}+\Gamma_{\circlearrowright}=0,
\end{split}
\end{equation}
as $\rm QD_u$ is coupled only to lead $u$, resulting in $J_{\rm{N}}^u = 0$. This gives $J_{\rm{Q}}^u=J_{\rm{E}}^u=\kappa\Gamma_{\circlearrowright}$.
However, QD$_{\rm b}$ is coupled to leads $l$ and $r$, allowing steady state (spin-polarized) particle flow between them. Conservation of total energy and particle currents at steady state, $\sum_{\lambda}J_{\rm{E}}^\lambda=0$ and $\sum_{\lambda} J_{\rm{N}}^\lambda=0$ gives: 
\begin{equation}
  J_{\rm{E}}^u=-J_{\rm{E}}^{lr}=\kappa\Gamma_{\circlearrowright}; \quad \quad  J_{\rm{N}}^l=-J_{\rm{N}}^r.
\end{equation}
The explicit form of $\Gamma_{\circlearrowright}$ is provided in Appendix-~\ref{Appendix-C}. Unlike energy, heat current is not conserved, as $J_{\rm{Q}}^{l}+J_{\rm{Q}}^{r}\equiv J_{\rm{Q}}^{lr}\neq J_{\rm{E}}^{lr}=-\kappa\Gamma_{\circlearrowright}$.

\subsection{Entropy Production Rate: Entropic flux \& bias}\label{Sec-V}

From Eq.~\eqref{S3}, one can  write~\cite{landi2021irreversible}
\begin{equation}\label{EP2}
\varDelta\mathcal{S}_{\rm{s}}(t)=\Sigma(t)+\Phi(t),
\end{equation}
where $\Sigma(t)$ is the \textit{entropy production} and $\Phi(t)$ is the \textit{entropy flux}. The explicit forms of $\Sigma(t)$ and the $\Phi(t)$ are derived following Refs.~\cite{strasberg2022quantum} [See Appendix-~\ref{Appendix-D}]:
\begin{equation}\label{EP3}
\begin{split}
\Sigma(t)=&k_B\Tr[\rho_{\rm{tot}}(t)\ln\{\rho_{\rm{tot}}(t)\}]\\
&-k_B\Tr[\rho_{\rm{tot}}(t)\ln\left\{\rho_{\rm{s}}(t)\prod_{\lambda}\rho_{\lambda}^{\rm{eq}}\right\}],\\
\Phi(t)=&k_B\sum_{\lambda}\Tr_{\rm{\lambda}}\left[\left\{\rho_{\lambda}(t)-\rho_{\lambda}^{\rm{eq}}\right\}\ln{\rho_{\lambda}^{\rm{eq}}}\right],
\end{split}
\end{equation}
where, $\rho_{\lambda}^{\rm{eq}}$ and $\rho_{\lambda}(t)$ are the equilibrium and \textit{near equilibrium} density operators of reservoir $\lambda$, respectively. In the grand canonical ensemble, $\rho_{\lambda}^{\rm{eq}}$ is expressed as:
\begin{equation}\label{EP4}
\rho_{\lambda}^{\rm{eq}}=\frac{e^{-\beta_{\lambda}H_{\rm{B}}^{\lambda}}}{\mathcal{Z_{\lambda}(\beta_{\lambda}})}, 
\end{equation}
where $H_{\rm{B}}^{\lambda}\equiv H_{\rm{B}}^{\lambda\sigma}=\sum_{k} (\epsilon^{\lambda\sigma}_{k}-\mu_{\lambda\sigma})c_{{\lambda\sigma} k}^\dagger c_{{\lambda\sigma} k}$ and ${\mathcal{Z_{\lambda}(\beta_{\lambda}})}=\Tr[e^{-\beta_{\lambda}H_{\rm{B}}^{\lambda}}]$ is the the grand canonical partition function. As a result, $\Phi(t)$ is identified as the reversible heat exchanged with the reservoir between some final and initial times
\begin{equation}\label{EP5}
\Phi(t)
=-k_B\sum_{\lambda}\beta_{\lambda}[{\langle H_{\rm{B}}^{\lambda}\rangle}_{t}-{\langle H_{\rm{B}}^{\lambda}\rangle}_{0}]
=k_B\sum_{\lambda}\beta_{\lambda}\varDelta Q_{\lambda}(t),
\end{equation}
where $\varDelta Q_{\lambda}(t)={\langle H_{\rm{B}}^{\lambda}\rangle}_{0}-{\langle H_{\rm{B}}^{\lambda}\rangle}_{t}$ and ${\langle H_{\rm{B}}^{\lambda}\rangle}_{t}=\Tr_{\rm{\lambda}}[\rho_{\lambda}(t)H_{\rm{B}}^{\lambda}]$. Thus, we rewrite Eq.~\eqref{EP2} as
\begin{equation}\label{EP6}
\varDelta\mathcal{S}_{\rm{s}}(t)=\Sigma(t)+k_B\sum_{\lambda}\beta_{\lambda}\varDelta Q_{\lambda}(t),
\end{equation}
where we replace the second term by Eq.~\eqref{EP5}. Therefore, the general definition of the entropy production rate can be deduced from the above equation as follows,
\begin{equation}\label{EP7}
\begin{split}
\dot{\Sigma}(t)&=\frac{d}{dt}\left\{\varDelta\mathcal{S}_{\rm{s}}(t)\right\}-k_B\sum_{\lambda}\beta_{\lambda}\frac{d}{dt}\left\{\varDelta Q_{\lambda}(t)\right\}\\
&=\frac{d}{dt}\left\{\varDelta\mathcal{S}_{\rm{s}}(t)\right\}-k_B\sum_{\lambda}\beta_{\lambda}J_{\rm{Q}}^\lambda(t).
\end{split}
\end{equation}

At steady state, the system's entropy remains constant, i.e., $\frac{d}{dt}\left\{\varDelta\mathcal{S}_{\rm{s}}(t)\right\}=0$, and $\dot{\Sigma}$ reaches its minimum~\cite{landi2021irreversible,strasberg2022quantum}. So, Eq.~\eqref{EP7} yields
\begin{equation}\label{EP8}
\dot{\Sigma}=-k_B\sum_{\lambda}\beta_{\lambda}J_{\rm{Q}}^\lambda.
\end{equation}
The above equation expresses the entropy production rate in terms of the appropriate thermodynamic fluxes and the corresponding temperatures of the various leads. Substituting $J_{\rm{Q}}^\lambda=J_{\rm{E}}^\lambda-\mu_{\lambda}J_{\rm{N}}^\lambda$ into Eq.~\eqref{EP8}, the entropy production rate for our three-terminal system can be expressed as
\begin{equation}\label{EP13}
\dot{\Sigma}=J_{\rm{E}}^u\mathcal{F}_{\rm{E}}^u+J_{\rm{E}}^r\mathcal{F}_{\rm{E}}^r+J_{\rm{N}}^r\mathcal{F}_{\rm{N}}^r,
\end{equation}
where, $J_{\rm{N}}^u=0$, since the upper QD is connected to only one lead (see also Eq.~\eqref{Jer-Jnr}). In deriving Eq.~\eqref{EP13}, we have also utilized the conservation laws for the energy and particle currents: $J_{\rm E}^{u}+ J_{\rm E}^{l}+J_{\rm E}^{r}= 0$ and $J_{\rm N}^{l}+J_{\rm N}^{r}= 0$. The conjugate ``forces" associated with the set of thermodynamic fluxes $\{J_{\rm{E}}^{u}$, $J_{\rm{E}}^{r}$,  $J_{\rm{N}}^r\}$ in Eq.~\eqref{EP13} are given by
\begin{equation}\label{EP14}
\begin{split}
\mathcal{F}_{\rm{E}}^{u} =& k_B(\beta_l-\beta_{u});\\
\mathcal{F}_{\rm{E}}^{r} =& k_B(\beta_l-\beta_{r});\\
\mathcal{F}_{\rm{N}}^r =& k_B(\beta_r\mu_r-\beta_l\mu_l).
\end{split}
\end{equation}
It is important to note that Eq.~\eqref{EP13} is not a unique representation of the entropy production rate, because the three energy fluxes are subject to one constraint arising from energy conservation. Consequently, one may choose any two independent energy fluxes (e.g., $\{J_{\rm{E}}^u,J_{\rm{E}}^r\}$ or $\{J_{\rm{E}}^u,J_{\rm{E}}^l\}$ or $\{J_{\rm{E}}^r,J_{\rm{E}}^l\}$) together with one particle flux (either $J_{\rm{N}}^r$ or $J_{\rm{N}}^l$), along with their respective conjugate \textit{forces} to write the entropy production rate. Therefore, there exist six equivalent representations of $\dot{\Sigma}$ that describe the same three-terminal system. The non-uniqueness in defining the entropic forces and corresponding fluxes does not signal any mathematical inconsistency; rather, it is an inherent feature of a general three-terminal transport. Although the form of energy biases depends on the choice of the force–flux representation, all such representations are thermodynamically equivalent and yield identical physical predictions across the different permissible formulations of entropy production rate. Therefore, without loss of generality, we adopt one particular form of $\dot{\Sigma}$ [cf. Eq.~\eqref{EP13}] throughout the remainder of this paper. If a different form of $\dot{\Sigma}$, is chosen, the parameter regimes for the occurrence of ICC or any other thermodynamic effects in the corresponding entropy production rate can be readily determined from the relation between $\{\mathcal{F}_{\rm{E}}^{u},\mathcal{F}_{\rm{E}}^{r},\mathcal{F}_{\rm{N}}^{r}\}$ in the present formulation [cf.~Eq.~\eqref{EP14}] and those in the alternative one. The quantities $\{\mathcal{F}_{\rm{E}}^{u},\mathcal{F}_{\rm{E}}^{r},\mathcal{F}_{\rm{N}}^{r}\}$ defined in Eqs.~\eqref{EP13}~and~\eqref{EP14} should not be confused with the \textit{real} thermal and particle forces acting on the system. They should instead be regarded as \textit{entropic} energy and particle biases (i.e., energy and particle biases in the entropy representation) that are conjugate to the complete set of individual fluxes $\{J_{\rm{E}}^u,J_{\rm{E}}^r,J_{\rm{N}}^r\}$. In the case of coupled transport involving a single pair of forces and fluxes, this correspondence becomes unique, and the entropic biases reduce to the real thermodynamic forces experienced by the system.

Before we analyze the coupled transport, let us first identify the microscopic descriptions of the above thermodynamic quantities. The ideal starting point could be the entropy production rate, which sets the connection between macroscopic and microscopic thermodynamic frameworks~\cite{schnakenberg1976network}. So, rewriting the von-Neumann entropy defined in Eq.~\eqref{S1} 
\begin{equation}\label{S-vn}
\mathcal{S}_{\rm{s}}(t)=-k_B \sum_{\mathbb{i}}\rho_{\mathbb{i}}(t)\ln{\rho_{\mathbb{i}}(t)},    
\end{equation}
in terms of the microscopic populations of the different system eigenstates, $\{\rho_{\mathbb{i}}\}$ ($\mathbb{i=A,B,C,D}$), we evaluate 
the time evolution of the system's entropy change as~\cite{schnakenberg1976network}
\begin{equation}\label{delta-s-dot}
\begin{split}
&\frac{d}{dt}\varDelta\mathcal{S}_{\rm{s}}(t)
=k_B\Bigg[\Gamma^{l+}_\mathbb{AB}\ln(\frac{\rho_{\mathbb{A}}}{\rho_{\mathbb{B}}})+\Gamma^{r+}_\mathbb{AB}\ln(\frac{\rho_{\mathbb{A}}}{\rho_{\mathbb{B}}})+\Gamma^{u+}_\mathbb{BD}\ln(\frac{\rho_{\mathbb{B}}}{\rho_{\mathbb{D}}})\\
&+\Gamma^{l-}_\mathbb{DC}\ln(\frac{\rho_{\mathbb{D}}}{\rho_{\mathbb{C}}})+\Gamma^{r-}_\mathbb{DC}\ln(\frac{\rho_{\mathbb{D}}}{\rho_{\mathbb{C}}})+\Gamma^{u-}_\mathbb{CA}\ln(\frac{\rho_{\mathbb{C}}}{\rho_{\mathbb{A}}})\Bigg],  
\end{split}
\end{equation}
where we use $\dot{\rho}_{\mathbb{i}}(t)$ from Eq.~\eqref{rho2}~[See Appendix-~\ref{Appendix-E}]. We emphasize that, since there are no off-diagonal (quantum coherence) elements in the reduced density matrix in Eqs.~\eqref{LME}–\eqref{Lindbladian}, the LME effectively reduces to a rate equation [cf.~Eq.~\eqref{rho2}] expressed in terms of occupation probabilities. As a result, the von Neumann entropy in Eq.~\eqref{S-vn} coincides with the Shannon entropy within its diagonalized description. Comparing the above equation with Eq.~\eqref{EP2}, we identify the microscopic version of the entropy production rate and entropy flux, as follows [Appendix-~\ref{Appendix-E}]
\begin{widetext}
\begin{equation}\label{sigma-phi-gen}
\begin{split}
\dot{\Sigma}(t)
=&k_B\Bigg[({\rm{k}}_\mathbb{AB}^{l+}\rho_{\mathbb{A}}-{\rm{k}}_\mathbb{BA}^{l-}\rho_{\mathbb{B}})\ln(\frac{{\rm{k}}_\mathbb{AB}^{l+}\rho_{\mathbb{A}}}{{\rm{k}}_\mathbb{BA}^{l-}\rho_{\mathbb{B}}})+({\rm{k}}_\mathbb{AB}^{r+}\rho_{\mathbb{A}}-{\rm{k}}_\mathbb{BA}^{r-}\rho_{\mathbb{B}})\ln(\frac{{\rm{k}}_\mathbb{AB}^{r+}\rho_{\mathbb{A}}}{{\rm{k}}_\mathbb{BA}^{r-}\rho_{\mathbb{B}}})+({\rm{k}}_\mathbb{BD}^{u+}\rho_{\mathbb{B}}-{\rm{k}}_\mathbb{DB}^{u-}\rho_{\mathbb{D}})\ln(\frac{{\rm{k}}_\mathbb{BD}^{u+}\rho_{\mathbb{B}}}{{\rm{k}}_\mathbb{DB}^{u-}\rho_{\mathbb{D}}})\\
+&({\rm{k}}_\mathbb{DC}^{l-}\rho_{\mathbb{D}}-{\rm{k}}_\mathbb{CD}^{l+}\rho_{\mathbb{C}})\ln(\frac{{\rm{k}}_\mathbb{DC}^{l-}\rho_{\mathbb{D}}}{{\rm{k}}_\mathbb{CD}^{l+}\rho_{\mathbb{C}}})+({\rm{k}}_\mathbb{DC}^{r-}\rho_{\mathbb{D}}-{\rm{k}}_\mathbb{CD}^{r+}\rho_{\mathbb{C}})\ln(\frac{{\rm{k}}_\mathbb{DC}^{r-}\rho_{\mathbb{D}}}{{\rm{k}}_\mathbb{CD}^{r+}\rho_{\mathbb{C}}})+({\rm{k}}_\mathbb{CA}^{u-}\rho_{\mathbb{C}}-{\rm{k}}_\mathbb{AC}^{u+}\rho_{\mathbb{A}})\ln(\frac{{\rm{k}}_\mathbb{CA}^{u-}\rho_{\mathbb{C}}}{{\rm{k}}_\mathbb{AC}^{u+}\rho_{\mathbb{A}}})\Bigg],\\
\dot{\Phi}(t)=&-k_B\Bigg[\Gamma^{l+}_\mathbb{AB}\ln(\frac{{\rm{k}}_\mathbb{AB}^{l+}}{{\rm{k}}_\mathbb{BA}^{l-}})+\Gamma^{r+}_\mathbb{AB}\ln(\frac{{\rm{k}}_\mathbb{AB}^{r+}}{{\rm{k}}_\mathbb{BA}^{r-}})+\Gamma^{u+}_\mathbb{BD}\ln(\frac{{\rm{k}}_\mathbb{BD}^{u+}}{{\rm{k}}_\mathbb{DB}^{u-}})
+\Gamma^{l-}_\mathbb{DC}\ln(\frac{{\rm{k}}_\mathbb{DC}^{l-}}{{\rm{k}}_\mathbb{CD}^{l+}})+\Gamma^{r-}_\mathbb{DC}\ln(\frac{{\rm{k}}_\mathbb{DC}^{r-}}{{\rm{k}}_\mathbb{CD}^{r+}})+\Gamma^{u-}_\mathbb{CA}\ln(\frac{{\rm{k}}_\mathbb{CA}^{u-}}{{\rm{k}}_\mathbb{AC}^{u+}})\Bigg].
\end{split}
\end{equation}
\end{widetext}
In Eq.~\eqref{sigma-phi-gen}, each term of the Schnakenberg entropy production rate $\dot{\Sigma}$~\cite{schnakenberg1976network}, has the form $(a-b)\ln(\frac{a}{b})$, ensuring non-negativity of the total entropy production rate. At steady state, this leads to [Appendix-~\ref{Appendix-E}]
\begin{widetext}
\begin{equation}\label{ss-sigma}
\begin{split}
\dot{\Sigma}(t)=-\dot{\Phi}(t)
=&\kappa\Gamma_{\circlearrowright} \Bigg[\left(\frac{k_B}{\kappa}\right)\ln(\frac{{\rm{k}}_\mathbb{AB}^{l+}{\rm{k}}_\mathbb{BD}^{u+}{\rm{k}}_\mathbb{DC}^{l-}{\rm{k}}_\mathbb{CA}^{u-}}{{\rm{k}}_\mathbb{BA}^{l-}{\rm{k}}_\mathbb{DB}^{u-}{\rm{k}}_\mathbb{CD}^{l+}{\rm{k}}_\mathbb{AC}^{u+}})\Bigg]
+\left(\varepsilon_{\rm{b}}\Gamma^{r+}_\mathbb{AB}-(\varepsilon_{\rm{b}}+\kappa)\Gamma^{r-}_\mathbb{DC}\right)\Bigg[\left(\frac{k_B}{\kappa}\right)\ln(\frac{{\rm{k}}_\mathbb{BA}^{r-}{\rm{k}}_\mathbb{AB}^{l+}{\rm{k}}_\mathbb{CD}^{r+}{\rm{k}}_\mathbb{DC}^{l-}}{{\rm{k}}_\mathbb{DC}^{r-}{\rm{k}}_\mathbb{CD}^{l+}{\rm{k}}_\mathbb{AB}^{r+}{\rm{k}}_\mathbb{BA}^{l-}})\Bigg]\\
+&(\Gamma^{r+}_\mathbb{AB}-\Gamma^{r-}_\mathbb{DC})\Bigg[k_B (1+\theta)\ln(\frac{{\rm{k}}_\mathbb{AB}^{r+}{\rm{k}}_\mathbb{BA}^{l-}}{{\rm{k}}_\mathbb{BA}^{r-}{\rm{k}}_\mathbb{AB}^{l+}})+k_B \theta\ln(\frac{{\rm{k}}_\mathbb{DC}^{r-}{\rm{k}}_\mathbb{CD}^{l+}}{{\rm{k}}_\mathbb{CD}^{r+}{\rm{k}}_\mathbb{DC}^{l-}})\Bigg] \equiv J_{\rm{E}}^u\mathcal{F}_{\rm{E}}^u+J_{\rm{E}}^r\mathcal{F}_{\rm{E}}^r+J_{\rm{N}}^r\mathcal{F}_{\rm{N}}^r,
\end{split}
\end{equation}    
\end{widetext}
where $\theta=(\frac{\varepsilon_{\rm{b}}}{\kappa})$ is the scaled system parameter. The sign of $\theta$ plays a crucial role in determining the ICC behavior in energy and (spin-polarized) particle currents, as discussed in Sec.~\ref{Sec-VIII}. From the above equation, all associated fluxes and their conjugate forces (entropic bias) are identified as follows:
\begin{equation}\label{flux}
\begin{split}
J_{\rm{E}}^{u}=&\kappa\Gamma_{\circlearrowright}\quad;\quad
J_{\rm{N}}^r=(\Gamma^{r+}_\mathbb{AB}+\Gamma^{r+}_\mathbb{CD}),\\
J_{\rm{E}}^r=&\varepsilon_{\rm{b}}\Gamma_\mathbb{AB}^{r+}+(\varepsilon_{\rm{b}}+\kappa)\Gamma_\mathbb{CD}^{r+}=\varepsilon_{\rm{b}}J_{\rm{N}}^r+\kappa\Gamma_\mathbb{CD}^{r+};
\end{split}
\end{equation}
and
\begin{subequations}\label{force}
\begin{align}   
\mathcal{F}_{\rm{E}}^{u}=&\left(\frac{k_B}{\kappa}\right)\ln(\frac{{\rm{k}}_\mathbb{AB}^{l+}{\rm{k}}_\mathbb{BD}^{u+}{\rm{k}}_\mathbb{DC}^{l-}{\rm{k}}_\mathbb{CA}^{u-}}{{\rm{k}}_\mathbb{BA}^{l-}{\rm{k}}_\mathbb{DB}^{u-}{\rm{k}}_\mathbb{CD}^{l+}{\rm{k}}_\mathbb{AC}^{u+}}) = k_B(\beta_l-\beta_{u}),\label{force-e-r}\\
\mathcal{F}_{\rm{E}}^{r}=&\left(\frac{k_B}{\kappa}\right)\ln(\frac{{\rm{k}}_\mathbb{BA}^{r-}{\rm{k}}_\mathbb{AB}^{l+}{\rm{k}}_\mathbb{CD}^{r+}{\rm{k}}_\mathbb{DC}^{l-}}{{\rm{k}}_\mathbb{DC}^{r-}{\rm{k}}_\mathbb{CD}^{l+}{\rm{k}}_\mathbb{AB}^{r+}{\rm{k}}_\mathbb{BA}^{l-}})\\
=&\left(\frac{k_B}{\kappa}\right)\ln\mathcal{\bigg(\frac{N}{M}\bigg)}=k_B(\beta_l-\beta_{r}),\label{force-e-b}\\
\mathcal{F}_{\rm{N}}^r=&k_B (1+\theta)\ln(\frac{{\rm{k}}_\mathbb{AB}^{r+}{\rm{k}}_\mathbb{BA}^{r-}}{{\rm{k}}_\mathbb{BA}^{l-}{\rm{k}}_\mathbb{AB}^{l+}})-k_B \theta\ln(\frac{{\rm{k}}_\mathbb{CD}^{r+}{\rm{k}}_\mathbb{DC}^{l-}}{{\rm{k}}_\mathbb{DC}^{r-}{\rm{k}}_\mathbb{CD}^{l+}})\nonumber\\
=&k_B (1+\theta)\ln{\mathcal{M}}-k_B \theta\ln{\mathcal{N}}\label{force-n-b-2}\\
=&k_B\ln(\mathcal{M})-\varepsilon_{\rm{b}}\mathcal{F}_{\rm{E}}^r=k_B(\beta_r\mu_r-\beta_l\mu_l)\label{force-n-b}
\end{align}
\end{subequations}
where we have defined two important factors as follows
\begin{equation}\label{MN}
\mathcal{M}=\left(\frac{{\rm{k}}_\mathbb{AB}^{r+}{\rm{k}}_\mathbb{BA}^{l-}}{{\rm{k}}_\mathbb{BA}^{r-}{\rm{k}}_\mathbb{AB}^{l+}}\right)\quad;\quad \mathcal{N}=\left(\frac{{\rm{k}}_\mathbb{CD}^{r+}{\rm{k}}_\mathbb{DC}^{l-}}{{\rm{k}}_\mathbb{DC}^{r-}{\rm{k}}_\mathbb{CD}^{l+}}\right)
\end{equation}

Now, Equation~\eqref{flux} is identical to Eq.~\eqref{EP14}. The equivalence between macroscopic and microscopic expressions of the above entropic forces [Cf.~Eq.~\eqref{force}] can be verified using Eq.~\eqref{ME15}, and the corresponding FDFs. Equations~\eqref{flux}-\eqref{force} are the first major results of our analysis that follow the linearity of the underlying quantum master equation [Cf.~Eq.~\eqref{rho2}]. Here, it is useful to highlight a point that may at first appear paradoxical. From Eq.~\eqref{force}, it might seem that although the microscopic expressions of the entropic biases depend on the sign of $\kappa$, their macroscopic expressions are independent of it. This arises because ${\mathrm{k}}_{\mathbb{ij}}^{\lambda \pm}$ (and consequently $\mathcal{M}$ and $\mathcal{N}$ in Eq.~\eqref{MN}) depend on the sign of $\kappa$ via $\omega_{\mathbb{ij}}$, which appears in FDF [cf.~Eq.~\eqref{ME15}]. Therefore, microscopically, depending on the sign of $\kappa$, the signs of ${\mathrm{k}}_{\mathbb{ij}}^{\lambda \pm}$ within the logarithm (or equivalently $\mathcal{M}$ and $\mathcal{N}$) adjust such that the same macroscopic entropic bias emerges, irrespective of the nature of $\kappa$. This freedom in the microscopic realization of the entropic biases allows us to establish the counterintuitive transport phenomenon of ICC in \textit{nanoscale quantum setting} with capacitively coupled QDs, as considered in the following section. In this precise microscopic sense, ICC in quantum systems is fundamentally different from the classical ICC model put forwarded by Wang et. al.~\cite{wang2020inverse}, where it originates from self-organization within the system in response to the applied force. However, from a thermodynamic perspective, the ICC phenomenon in both classical and quantum systems exhibits the same richness: one induced current must flow against two mutually parallel thermodynamic forces without violating the laws of thermodynamics.

\section{Investigating Coupled transport}\label{Sec-VI}

Now, there are two effective approaches to establish our three terminal setup for a coupled transport with one energy and one particle force. The third possibility, in which both energy forces are present, could lead to anomalous energy flow; this has been studied in recent work~\cite{guan2025anamalous} and is therefore not discussed in this paper.

\subsection{$\beta_l=\beta_r=\beta$: QD Thermoelectric Setup}

\begin{figure}[t]
\centering    
\includegraphics[width=\columnwidth,height=3.9cm]{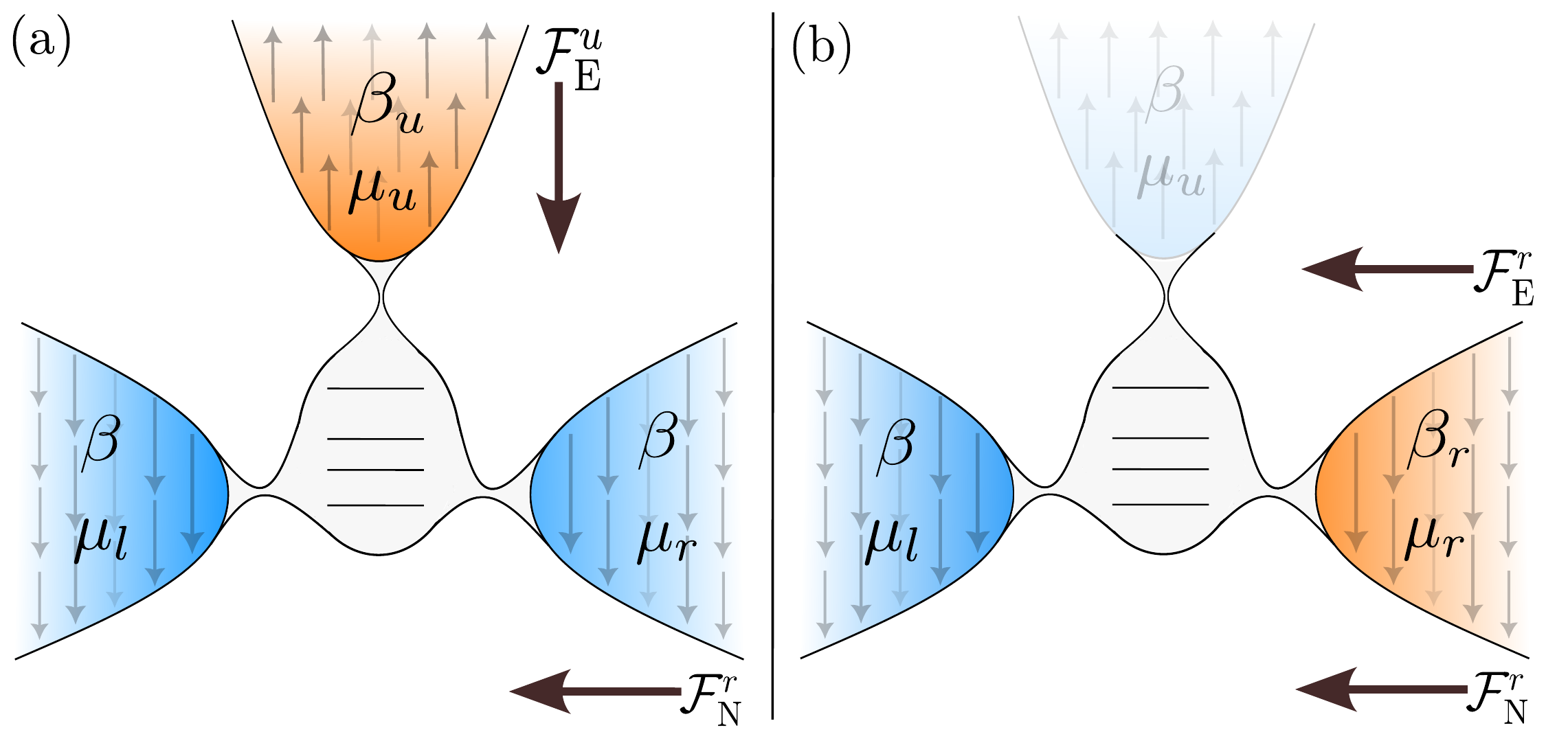}
\caption{(a) Coupled transport phenomena in QD thermoelectric system. A three-terminal energy harvester based on capacitively coupled QDs in a 2DEG setup, experimentally realized by Thierschmann \textit{et al.}~\cite{thierschmann2015three}, where a thermal bias between the upper and lower terminals drives a current against a chemical potential bias. (b) QD ICC effect as a new kind of coupled transport phenomenon, where both thermodynamic forces are mutually parallel to each other, in contrast to the QD thermoelectric setup shown in Fig.~\ref{Fig.3}a.}
\label{Fig.3}
\end{figure}

Once we set $\beta_l = \beta_r = \beta$, we can immediately identify one energy force, $\mathcal{F}_{\rm E}^{u}$, and one particle force, $\mathcal{F}_{\rm N}^{r}$, operating exactly as shown in Fig.~\ref{Fig.3}a. This is already evident from Eq.~\eqref{force}. Under the above choice of parameters, one of the entropic forces, $\mathcal{F}_{\rm E}^{r}$, becomes zero, and the remaining two entropic biases unambiguously reduce to the energy force
$\mathcal{F}_{\rm E}^{u} = k_B(\beta - \beta_u)$
and the particle force
$\mathcal{F}_{\rm N}^{r} = k_B \beta (\mu_r - \mu_l)$, respectively, as depicted in Fig.~\ref{Fig.3}a. This configuration is equivalent to the S\'{a}nchez--B\"{u}ttiker model~\cite{sanchez2011optimal}, which is widely used to study thermoelectric effects in coupled QD systems~\cite{sanchez2011optimal,donsa2014double,thierschmann2015three,erdman2017thermoelectric,whitney2018quantum,wang2022cycleflux}, as well as various other quantum-transport and thermal-rectification phenomena~\cite{ruokola2011single,zhang2017three,aligia2020heat,tesser2022heat,shuvadip2022univarsal,yang2019thermal,kutvonen2016thermodynamics}.

In this limit, Eq.~\eqref{EP13} simplifies to
\begin{equation}\label{sigma-SB-model}
\dot{\Sigma} = J_{\rm E}^{u} \mathcal{F}_{\rm E}^{u} + J_{\rm N}^{r} \mathcal{F}_{\rm N}^{r}.   
\end{equation}
It is important to emphasize that although Eq.~\eqref{EP13} is not, in general, a unique representation of the entropy production rate, once one of the entropic biases becomes zero (here, $\mathcal{F}_{\rm E}^{r}=0$), Eq.~\eqref{EP13} describes coupled transport for physical systems with well-defined thermodynamic forces.

However, the configuration shown in Fig.~\ref{Fig.3}a cannot exhibit ICC, because an induced current cannot simultaneously flow against both thermodynamic forces, $\mathcal{F}_{\rm E}^{u}$ and $\mathcal{F}_{\rm N}^{r}$, as they are not \textit{mutually parallel}. Nevertheless, the setup in Fig.~\ref{Fig.3}a can still operate as a thermoelectric engine or as a refrigerator. For example, when a particle current $J_{\rm N}^r$, induced by the energy force $\mathcal{F}_{\rm E}^{u}$, flows against the particle force $\mathcal{F}_{\rm N}^{r}$, the device functions as an engine. In contrast, when an energy current $J_{\rm E}^{u}$, induced by $\mathcal{F}_{\rm N}^{r}$, flows against its conjugate force $\mathcal{F}_{\rm E}^{u}$, the device acts as a refrigerator. This behavior relies on the well-known principle of thermodynamic cross-effects~\cite{sanchez2011optimal,thierschmann2015three,wang2022cycleflux,gupt2024graph}, where an applied force drives a current that can oppose its conjugate force without violating the positivity of the total entropy production rate. An analogous model, shown in Fig.~\ref{Fig.3}a, was theoretically proposed by S\'{a}nchez and B\"{u}ttiker~\cite{sanchez2011optimal} and experimentally realized by Thierschmann \textit{et al.}~\cite{thierschmann2015three} as a three-terminal energy harvester based on capacitively coupled QDs in a 2DEG setup. In the experiment, a thermal bias of approximately $100\,\mathrm{mK}$ is applied between the upper and lower terminals. As a result, a current of the order of pA is measured in the lower terminals against a chemical potential bias of approximately $10\,\mu\mathrm{V}$. An extension of this model in the presence of spin-bias voltages has also been proposed by the present authors~\cite{gupt2024graph}.

In contrast to the present case --- where an induced current can oppose only one thermodynamic force at a time (for e.g, in a thermoelectric refrigerator, $J_{\rm E}^{u}$ opposes $\mathcal{F}_{\rm E}^{u}$ while $J_{\rm N}^{r}$ still aligns with $\mathcal{F}_{\rm N}^{r}$, otherwise they will violate the 2nd law)—the configuration in Fig.~\ref{Fig.3}b features mutually parallel thermodynamic forces. In that scenario, a current can flow against both forces while still satisfying $\dot{\Sigma} \ge 0$. This is the hallmark of the ICC effect proposed by Wang \textit{et al.}~\cite{wang2020inverse} in classical settings.

\subsection{$\beta_l=\beta_u=\beta$: QD ICC effect} 

If we set $\beta_l = \beta_u = \beta$, this is equivalent to imposing that the other entropic energy bias $\mathcal{F}_{\rm E}^{u}$ is identically zero. Under this choice, Eq.~\eqref{EP8} simplifies to  
\begin{equation}\label{sigma-ICC}
\dot{\Sigma} = J_{\rm E}^{r}\,\mathcal{F}_{\rm E}^{r} + J_{\rm N}^{r}\,\mathcal{F}_{\rm N}^{r} \quad \geq  \quad 0,
\end{equation}
where the two real \textit{thermodynamic forces}, are given by $\mathcal{F}_{\rm E}^{r} = k_{\mathrm B}(\beta - \beta_r)$ and $\mathcal{F}_{\rm N}^{r} = k_{\mathrm B}(\beta_r \mu_r - \beta \mu_l)$, respectively [Fig.~\ref{Fig.3}b]. Without loss of any generality, if they are \textit{mutually parallel} with $\mathcal{F}_{\rm E}^{r}>0$ and $\mathcal{F}_{\rm N}^{r} >0$, either $J_{\rm E}^{r}$ or $J_{\rm N}^{r}$ can flow opposite to both forces while the second flux compensates for the resulting negative contribution, ensuring that the total entropy production still satisfies $\dot{\Sigma} \ge 0$. As a result, Fig.~\ref{Fig.3}b represents the ideal configuration for achieving ICC as a distinct thermodynamic phenomenon: a regime where one current can flow against two mutually parallel thermodynamic forces, without violating the laws of thermodynamics.

\begin{figure}[b]
    \centering   \includegraphics[width=0.95\columnwidth,height=9cm]{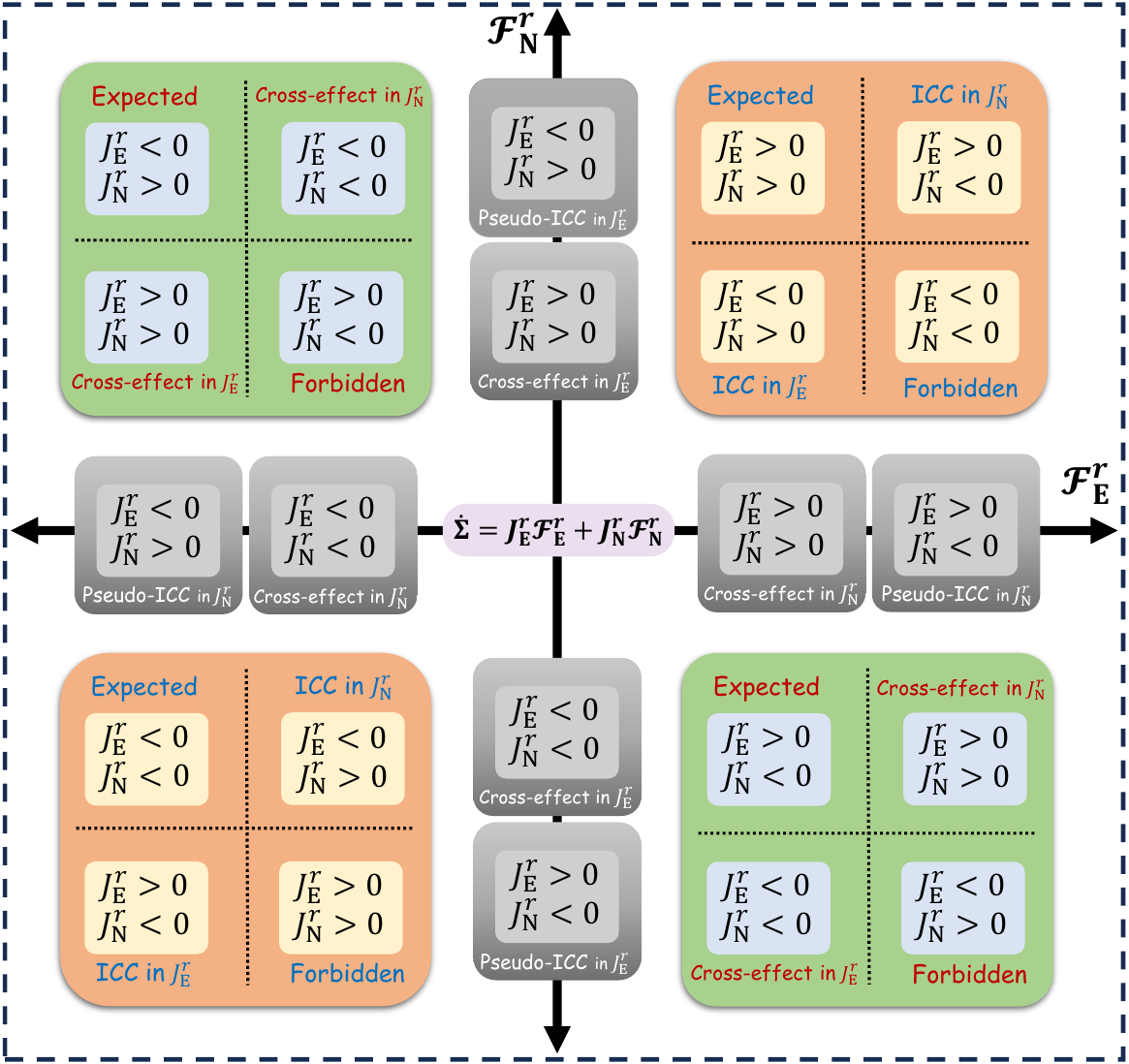}
    \caption{Pictorial representation of various thermodynamic phenomena governed by the entropy production rate $\dot{\Sigma} =J_{\rm{E}}^{r}\mathcal{F}_{\rm{E}}^{r}+J_{\rm{N}}^{r}\mathcal{F}_{\rm{N}}^{r}$ [cf.~Eq.~\eqref{sigma-ICC}] i.e., normal cross-effect (in green shaded region), ICC (orange shaded region), pseudo-ICC, Seebeck and Peltier effects (gray shaded blocks along $X$ and $Y$ axis) that can be obtained.}
    \label{ICC-chart}
\end{figure}

A special case of ICC is \emph{pseudo-ICC}, which occurs when one of the forces is set to zero. For example, if $\mathcal{F}_{\rm E}^{r}=0$, then the induced current $J_{\rm E}^{r}$ may flow along or against $\mathcal{F}_{\rm N}^{r}$ without affecting the nonnegativity of the entropy production rate, i.e., $\dot{\Sigma} = J_{\rm N}^{r}\,\mathcal{F}_{\rm N}^{r} \; \geq  \; 0$ [cf.~Eq.~\eqref{sigma-ICC}]. In such a situation, if $J_{\rm E}^{r}$ opposes $\mathcal{F}_{\rm N}^{r}$, it signals ICC, and we refer to it as ``pseudo-ICC" --- a precursor to genuine ICC, which is in complete accordance with second law of thermodynamics [See also Sec.~\ref{Sec-VIII}:Pesudo-ICC]. Following Wang \textit{et al.}~\cite{wang2020inverse}, here, we must distinguish ICC from normal cross-effects. For example, ICC in the energy (particle) current is realized by $J_{\rm E(N)}^{r}<0$ when both $\mathcal{F}_{\rm E}^{r}>0$ and $\mathcal{F}_{\rm N}^{r}>0$ [see Fig.~\ref{ICC-chart}]. In contrast, for a conventional cross-effect, $\mathcal{F}_{\rm E}^{r}$ and $\mathcal{F}_{\rm N}^{r}$ have opposite signs, i.e., the two forces oppose each other. Then the cross-effect in the energy current (characterized by $J_{\rm E}^{r}<0$) is induced by the particle force (characterized by $\mathcal{F}_{\rm N}^{r}<0$), flow against the energy force $\mathcal{F}_{\rm E}^{r}>0$ [see Fig.~\ref{ICC-chart}]. The same is true for the particle current with $J_{\rm N}^{r}<0$, when $\mathcal{F}_{\rm E}^{r}<0$ but $\mathcal{F}_{\rm N}^{r}>0$ [see Fig.~\ref{ICC-chart}].

In summary, while Fig.~\ref{Fig.3}a illustrates the conventional thermoelectric effect, Fig.~\ref{Fig.3}b captures the essence of ICC, characterized by mutually parallel thermodynamic forces [Cf.~Eq.~\eqref{sigma-ICC}]. Moreover, any alternative formulation of the entropy production rate involving two mutually parallel force-flux pairs is isomorphic to Fig.~\ref{Fig.3}b and Eq.~\eqref{sigma-ICC}. Thus, we may drop (for convenience of the later sections, we still keep) the superscripts in~\eqref{sigma-ICC},
while still capturing all possible behaviors in the $(\mathcal{F}_{\rm E}, \mathcal{F}_{\rm N})$ plane, as summarized schematically in Fig.~\ref{ICC-chart}. As shown there, ICC occurs in the first and third quadrants, whereas cross-effects arise in the second and fourth quadrants.

In the following section, we analyze this reduced two-force-flux setup as a prototype system to realize ICC. For all numerical plots, we set the tunnelling rates equal, $\gamma_l=\gamma_r=\gamma_u\equiv\gamma$, and employ appropriately scaled dimensionless forces and fluxes. All fluxes are expressed in units of $\hbar\gamma^2$, while the energy and particle forces are measured in units of $k_B/(\hbar\gamma)$ and $k_B$, respectively. Since the ICC appears only in the first and third quadrants [Fig.~\ref{ICC-chart}], in the following we may, without loss of generality, choose the scaled forces to be non-negative and restrict our discussion to the first quadrant.

\section{Examining ICC: Results and Discussion}~\label{Sec-VIII}

As discussed in the previous section, considering $\beta_l=\beta_u=\beta$, we set the entropic bias $\mathcal{F}_{\rm{E}}^{u}$ to zero.  Now, from Eq.~\eqref{flux}, it is clear that to determine the currents $J_{\rm{N}}^r=(\Gamma^{r+}_\mathbb{AB}+\Gamma^{r+}_\mathbb{CD})$ and $J_{\rm{E}}^r=\varepsilon_{\rm{b}}J_{\rm{N}}^r+\kappa\Gamma_\mathbb{CD}^{r+}$, we need to understand the signs of the rates $\Gamma_\mathbb{AB}^{r+}$ and $\Gamma_\mathbb{CD}^{r+}$. To this end, we define two variables: 
\begin{equation}\label{R7}
\mathcal{X}:\stackrel{\text{def.}}=\Gamma_\mathbb{AB}^{r+}-\Gamma_\mathbb{AB}^{l+};\quad\quad\mathcal{Y}:\stackrel{\text{def.}}=\Gamma_\mathbb{CD}^{r+}-\Gamma_\mathbb{CD}^{l+}.
\end{equation}
In terms of which the rates are given by
\begin{equation}\label{R8}
\Gamma_\mathbb{AB}^{r+}=\frac{1}{2}(\mathcal{X}+\Gamma_{\circlearrowright});\quad\quad 
\Gamma_\mathbb{CD}^{r+}=\frac{1}{2}(\mathcal{Y}-\Gamma_{\circlearrowright}),
\end{equation}
and the currents are given by
\begin{equation}
J_{\rm N}^r = \frac{1}{2}(\mathcal{X} + \mathcal{Y}); \quad J_{\rm{E}}^r=\frac{\varepsilon_{\rm{b}}}{2}(\mathcal{X} + \mathcal{Y})+\frac{\kappa}{2}(\mathcal{Y}-\Gamma_{\circlearrowright}).
\end{equation}
Thus, by simply knowing the signs of $\mathcal{X}$, $\mathcal{Y}$, and $\Gamma_{\circlearrowright}$ we can determine the sign of the particle current $J_{\rm{N}}^r$ and the energy current $J_{\rm{E}}^r$. Now, the variables $\mathcal{X,Y}$ are directly related to $\mathcal{M,N}$ of Eq.~\eqref{MN}. It is straightforward to show that $\mathcal{M}\gtrless1,\mathcal{N}\gtrless1$ corresponds to $\mathcal{X}\gtrless0$ and $\mathcal{Y}\gtrless0$ respectively, whereas $\mathcal{M}=\mathcal{N}=1$ implies $\mathcal{X}=\mathcal{Y}=0$. The determination of the sign of $\Gamma_{\circlearrowright}$ is little more involved and we refer to Appendix~\ref{Appendix-F} for details. This is our general recipe for determining the signs of both energy and particle currents. As a simple consistency check of the above procedure, we have shown in Appendix~\ref{Appendix-F}, when both forces are zero, both currents vanish, as expected.

\subsection{Pseudo-ICC}

Now, we analyze pseudo-ICC in the energy and particle currents setting 
$\mathcal{F}_{\rm E}^r = 0$ and $\mathcal{F}_{\rm N}^r = 0$, respectively, since pseudo-ICC serves as a precursor to genuine ICC.

\subsubsection{Setting $\mathcal{F}_{\rm{E}}^{r}=0$: Pseudo-ICC in $J_{\rm{E}}^r$}

\begin{figure}[t]
    \centering    \includegraphics[width=\columnwidth,height=3.8cm]{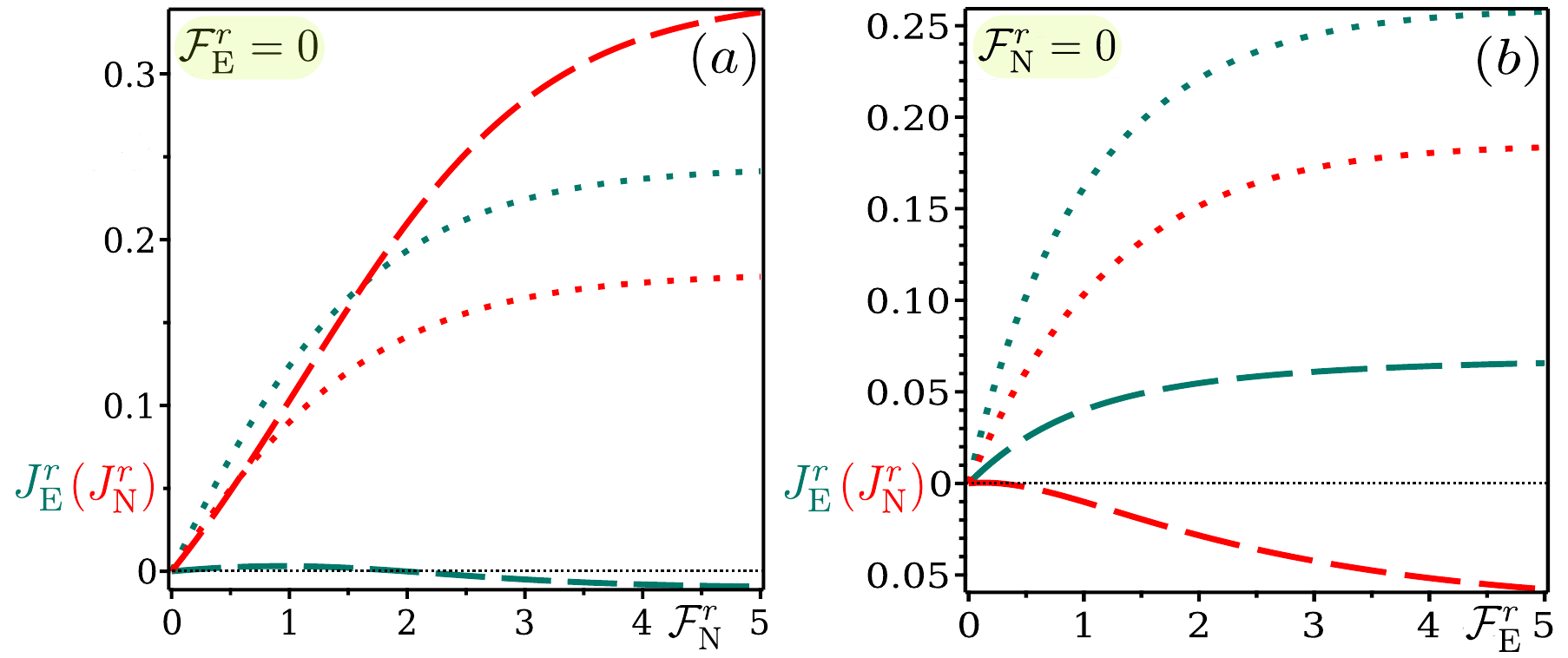}
    \caption{Pseudo-ICC in $J_{\rm{E}}^r$ (bluegreen line) and $J_{\rm{N}}^r$ (red line) against the thermodynamic forces (a) $\mathcal{F}_{\rm{N}}^{r}$ while $\mathcal{F}_{\rm{E}}^{r}=0$ and (b) $\mathcal{F}_{\rm{E}}^{r}$ while $\mathcal{F}_{\rm{N}}^{r}=0$, respectively. For all cases, the (dashed) dotted lines correspond to $\kappa=+1.5\hbar\gamma$ ($\kappa=-1.5\hbar\gamma$), validating the condition $|\kappa|>\varepsilon_{\rm{b}}$. Other system and bath parameters: $\varepsilon_{\rm{b}}=1.0\hbar\gamma$, $\varepsilon_{\rm{u}}=2.5\hbar\gamma$, $\beta_{r}=1/\hbar\gamma$, $\mu_{r}=1.0\hbar\gamma$.}
    \label{RM2}
\end{figure}

When $\mathcal{F}_{\rm N}^r$ is the only non-zero force, the  spin-polarized particle current $J_{\rm N}^r$ is always positive; while the energy current $J_{\rm E}^r$ remains positive for $\kappa > 0$, but can change sign when $\varepsilon_{\rm b} + \kappa < 0$ [Appendix~\ref{Appendix-F}]. A positive $J_{\rm E}^r$ corresponds to a cross effect, whereas a negative $J_{\rm E}^r$ signals pseudo-ICC, since the current flows against the non-conjugate particle force $\mathcal{F}_{\rm N}^r$. 
Thus, genuine ICC can arise in the energy current within this model [FIG.~\ref{RM2}(a)].

\subsubsection{Setting $\mathcal{F}_{\rm{N}}^{r}=0$: Pseudo-ICC in $J_{\rm{N}}^r$}

When $\mathcal{F}_{\rm E}^r$ is the only driving force, the energy current 
$J_{\rm E}^r$ remains positive [FIG.~\ref{RM2}(b)]. For $\kappa>0$, the particle current 
$J_{\rm N}^r$ is also positive, indicating no ICC. However, when $\varepsilon_{\rm b}+\kappa < 0$, 
$J_{\rm N}^r$ may become positive or negative [Appendix~\ref{Appendix-F}], enabling pseudo-ICC in the spin-polarized particle 
current [FIG.~\ref{RM2}(b)].

It follows from the above discussion that the occurrence of negative cross effects provides a strong indication for the emergence of a genuine ICC. As a result, an unconventional Seebeck or unconventional Peltier effect where an induced current flows against the existing thermodynamic force can be regarded as manifestations of pseudo-ICC (see Ref.~\cite{ghosh2026thermodynamics} for further details). Nevertheless, this scenario does not correspond to a true or genuine ICC, since the conjugate force attributed to the current is absent. In this sense, the pseudo-ICC regime serves as a precursor for the realization of genuine ICC behavior. Achieving genuine ICC is therefore considerably more challenging than observing pseudo-ICC, making it a particularly counterintuitive and nontrivial manifestation of coupled transport phenomena.

\subsection{Genuine ICC: When both forces are non-zero}
\par From the previous two cases, pseudo-ICC in both energy and particle currents occurs only when 
$\varepsilon_{\rm b}+\kappa < 0$, i.e., with negative $\kappa$. Therefore, the search for genuine ICC is restricted to the regime $-\kappa > \varepsilon_{\rm b} > 0$ (equivalently $-1 < \theta < 0$). Also, without loss of generality --- we focus only on the first quadrant of Fig.~\ref{ICC-chart}, where both forces $(\mathcal{F}_{\rm E}^r,\mathcal{F}_{\rm N}^r)$ are positive and mutually parallel. Because $\kappa<0$, Eq.~\eqref{force-e-b} implies $\mathcal{M}>\mathcal{N}$. The positivity of the particle force $\mathcal{F}_{\rm N}^r$ further requires $\mathcal{M}>1$, which in turn yields $\mathcal{X}>0$. Imposing positivity of both forces and using Eq.~\eqref{force-n-b-2}, we obtain
\begin{equation}
(\varepsilon_{\rm b}+\kappa)\ln \mathcal{M} < \varepsilon_{\rm b}\ln \mathcal{N},
\end{equation}
which allows $\mathcal{N}$ to be either greater or less than unity, i.e., $\mathcal{N}\gtrless 1$, corresponding to $\mathcal{Y}\gtrless 0$. We therefore arrive at two possible subcases: $\textit{1.}~\mathcal{M}>1,\, \mathcal{N}>1$, and $\textit{2.}~\mathcal{M}>1,\, \mathcal{N}<1$, which ultimately determine the conditions for achieving genuine ICC in $J_{\rm E}^r$ and $J_{\rm N}^r$, respectively.

\begin{figure}[b]
\includegraphics[width=0.84\columnwidth,height=12.4cm]{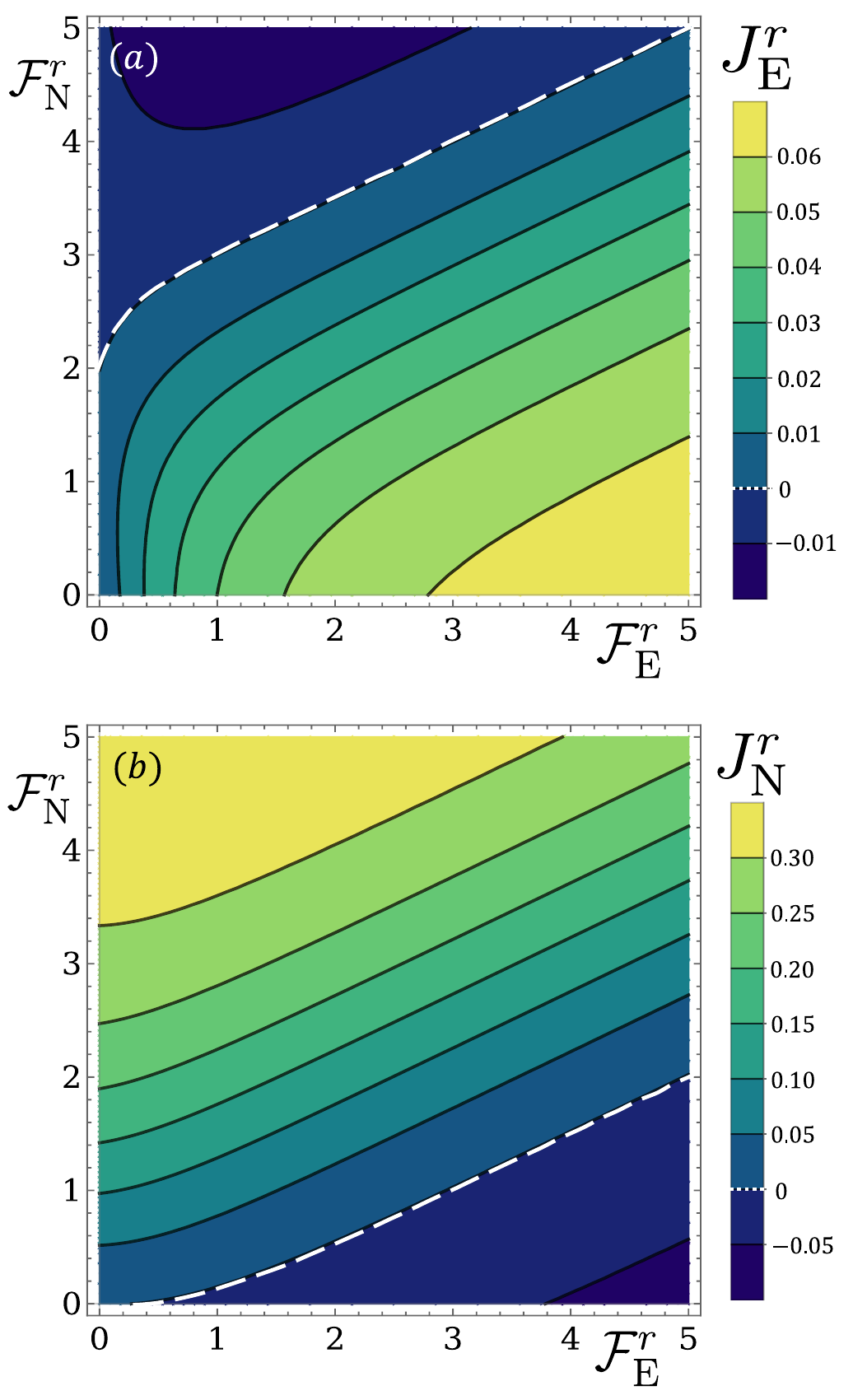}
\caption{Variation of (a) the spin-polarised particle current $J_{\rm{E}}^r$ and (b) the energy current $J_{\rm{N}}^r$ with both the thermodynamic forces for $\kappa=-1.5\hbar\gamma$. Genuine ICC for $J_{\rm{E}}^r$ occurs in the area above the white dashed line in (a) while it occurs for $J_{\rm{N}}^r$ below the white dashed line in (b). Other system and bath parameters: $\varepsilon_{\rm{b}}=1.0\hbar\gamma$, $\varepsilon_{\rm{u}}=2.5\hbar\gamma$, $\beta_{r}=1/\hbar\gamma$, $\mu_{r}=1.0\hbar\gamma$.}
\label{Contour2}
\end{figure}

\subsubsection{ICC in energy current $J_{\rm{E}}^r$}

Since both $\mathcal{M},\mathcal{N}>1$ (equivalently $\mathcal{X},\mathcal{Y}>0$), it follows directly that $J_{\rm N}^r=\tfrac{1}{2}(\mathcal{X}+\mathcal{Y})>0$. For $\kappa<0$ and $\mathcal{F}_{\rm E}^r>0$, we have
$\Gamma_{\circlearrowright}>0$, 
which implies the sign of $J_{\rm E}^r$ is not guaranteed, and the energy current may take either sign, $J_{\rm E}^r\gtrless 0$. Genuine ICC appears precisely when $J_{\rm E}^r<0$ [Fig.~\ref{Contour2}(a)], indicating energy flow against both positive forces $\mathcal{F}_{\rm N}^r$ and $\mathcal{F}_{\rm E}^r$. Thus, our present model can function as an autonomous thermoelectric refrigerator~\cite{zhang2023inverse}, driving energy flux against thermal gradient with a COP given by
\begin{equation}\label{cop2}
\zeta=\frac{-J_{\rm{E}}^r\cdot\beta_r}{J_{\rm{N}}^r\cdot\mathcal{F}_{\rm N}^{r}}=\zeta_R\left(1-\frac{\dot{\Sigma}}{J_{\rm{N}}^r\cdot\mathcal{F}_{\rm N}^{r}}\right)\leq \zeta_R,
\end{equation}
where $\zeta_R$ is the COP of an ideal refrigerator, defined as, $\zeta_R=T_{\mathrm{cold}}/(T_{\mathrm{hot}}-T_{\mathrm{cold}})$.

\subsubsection{ICC in particle current $J_{\rm{N}}^r$}

In the second case, defined by $\mathcal{M}>1$ and $\mathcal{N}<1$ (i.e., $\mathcal{X}>0$ and $\mathcal{Y}<0$), the particle current $J_{\rm N}^r=\tfrac{1}{2}(\mathcal{X}+\mathcal{Y})$
can be either positive or negative, $J_{\rm N}^r \gtrless 0$. Genuine ICC occurs when $J_{\rm N}^r<0$ [Fig.~\ref{Contour2}(b)], corresponding to particle flow against both positive forces $\mathcal{F}_{\rm N}^r$ and $\mathcal{F}_{\rm E}^r$. 

Thus, our reduced three-terminal model can be implemented as an autonomous (spin-thermoelectric) heat engine~\cite{benenti2022autonomous}, driving spin-polarized particle flux against the particle force, under the influence of the non-conjugate energy force $\mathcal{F}_{\rm{E}}^{r}$. The output power is defined as $|(J_{\rm{N}}^r\mathcal{F}_{\rm{N}}^{r}/\beta)|$, and the device efficiency is given by 
\begin{equation}\label{rm19}
\eta=\frac{-J_{\rm{N}}^r\cdot\mathcal{F}_{\rm{N}}^{r}}{\beta\cdot J_{\rm{E}}^r}=\eta_C-\frac{\dot{\Sigma}}{\beta\cdot J_{\rm{E}}^r}\leq \eta_C,
\end{equation}
where, $\eta_C=(T_{\rm{hot}}-T_{\rm{cold}})/T_{\rm{hot}}$.

To summarize, a genuine ICC is achieved in both energy-and spin-polarized particle currents when the two \textit{mutually parallel} forces $\mathcal{F}_{\rm{E}}^{r}$ and $\mathcal{F}_{\rm{N}}^{r}$ are positive. This allows the model to function as both a spin-thermoelectric heat engine and a refrigerator. The key takeaway from the present model is that the condition $-\kappa>\varepsilon_{\rm{b}}>0$ is essential for observing ICC in energy and particle currents, representing the swapping of energy levels between $\ket{0\uparrow}$($|\mathbb{C}\rangle$) and $\ket{\downarrow\uparrow}$($|\mathbb{D}\rangle$). For $\kappa>0$, the eigenstates are arranged in such a way [FIG.\ref{Model}(b)] that the particle (de)excitation implies energy (de)excitation. When  $-\kappa>\varepsilon_{\rm{b}}>0$, however, the reordering of the eigen-states [FIG.\ref{Model}(c)] creates an asymmetry between the particle and energy exchange, namely, the particle excitation between $\ket{0\uparrow}\rightarrow\ket{\downarrow\uparrow}$ implies the energetic de-excitation and vice-versa. It is worth mentioning that Wang \textit{et al.}~\cite{wang2020inverse} also identified analogous symmetry-breaking as a guiding factor for the emergence of ICC in classical Hamiltonian systems. Thus, it is intuitively evident that some form of symmetry-breaking is necessary to achieve inverse current in coupled quantum transport. In the present model with double QDs, an attractive interdot ($\kappa<0$) interaction~\cite{hamo2016electron,tabatabaei2018charge} serves as the necessary criterion, while $-\kappa>\varepsilon_{\rm{b}}>0$ acts as the sufficient condition for ICC by breaking the energy-particle exchange symmetry. Furthermore, the present results validate the second law of thermodynamics by ensuring that the two ICC regions do not overlap [FIG.~\ref{Contour2}]. Specifically, ICC in $J_{\rm{N}}^r$ occurs when both $\mathcal{M},\mathcal{N}$ are positive, whereas for $J_{\rm{E}}^r$, $\mathcal{N}$ must be negative, in accordance with the sufficient criteria for genuine ICC. Thus, our analysis highlights ICC as a counterintuitive thermodynamic phenomenon distinct from normal cross-effect, and particularly significant for coupled transport with two \textit{mutually parallel} thermodynamic forces.

\section{Conclusion}\label{Sec-IX}

To conclude, identifying the correct entropic bias and its conjugate fluxes in systems with multiple reservoirs is crucial for comprehending complex thermodynamic phenomena. Reducing the three-terminal model to a two-force transport system clarifies the distinction between ICC and normal cross-effects, since the three-terminal case, with its \textit{nonphysical} entropic bias, obscures the unambiguous identification of ICC. Above all, the present work develops a comprehensive quantum thermodynamic framework for the ICC, where one induced current flows counter to two mutually parallel thermodynamic forces in the system. Using a simple model of Coulomb-coupled QDs, we investigate the intriguing inverse current behavior in both energy and spin-polarized particle currents under near-equilibrium conditions. Our study examines the entropy production rate at both macroscopic and microscopic levels, leveraging the grand-canonical formalism of the Lindblad master equation and Schnakenberg's entropy formulation. The linear structure of the quantum master equation allows for exact analytical expressions for entropic bias and entropic fluxes, incorporating macroscopic reservoir parameters and microscopic system details. This approach uniquely identifies all possible entropic force-flux pairs for both general and coupled transport systems with \textit{mutually parallel} forces, enabling a systematic exploration of genuine ICC phenomena in energy and spin-polarized particle currents.

This could open potential applications for unconventional autonomous nano-thermoelectric engines and refrigerators based on the ICC effect in coupled QD systems. Notably, our analysis indicates that an autonomous QD refrigerator or engine requires attractive interactions between QDs as a necessary criterion~\cite{hamo2016electron,tabatabaei2018charge,little1964possibility,prawiroatmodjo2017negativeU} and highlights the possibility for more versatile spin-thermoelectric devices compared to traditional designs, with the novelty rooted in the counterintuitive thermodynamic behavior of inverse current. A similar concept for an autonomous circular heat engine, inspired by the ICC effect in classical systems~\cite{wang2020inverse}, has already been recently proposed by Benenti \textit{et al.}~\cite{benenti2022autonomous}. We also believe our findings will play a pivotal role in advancing ICC-assisted unconventional autonomous quantum thermal devices in the near future.

\section*{Acknowledgments}
AG acknowledges IITK for infrastructure and financial support. S.G. acknowledges the Ministry of Education, Government of India, for the Prime Minister Research Fellowship (PMRF). N.G. is grateful for the FARE fellowship from the IIT Kanpur.

\onecolumngrid 
\appendix

\section{Derivation of the Lindblad Master Equation}\label{Appendix-A}

\par We are going to derive the master equation in the interaction picture for studying the dynamical evolution of the system due to interaction with the reservoir through the exchange of both particles and energy. Let us start with the tunneling Hamiltonian, which essentially signifies the interaction,
\begin{equation}\label{ME1}
\begin{split}
  H_{\rm{T}}=H_{\rm{T}}^{{\rm{b}\downarrow}{l}}+H_{\rm{T}}^{{\rm{b}\downarrow}{r}}+H_{\rm{T}}^{{\rm{u}\uparrow}{u}}
  \end{split}
\end{equation}
where, $H_{\rm{T}}^{{\rm{b}\downarrow}{l(r)}}$ signifies the interaction of $\rm{QD_b}$ with the reservoir $l(r)$ through the exchange of spin-down electrons and similarly $H_{\rm{T}}^{{\rm{u}}\uparrow u}$ signifies the interaction of $\rm{QD_u}$ with the reservoir $u$ through the exchange of spin-up electrons, that are given by in the ~\eqref{HT} of the main text. To derive the master equation, we start with the von Neumann equation for the total density matrix $\rho_{\rm{tot}}$ in the interaction picture
\begin{equation}\label{A2}
    \frac{d}{dt}\rho_{\rm{tot}}=-\frac{i}{\hbar}[H_{\rm{T}}(t),\rho_{\rm{tot}}(t)].
\end{equation}
Integrating the above equation and	taking a trace over the bath degrees of freedom, one obtains
\begin{equation}\begin{split}\label{A3}
\frac{\partial}{\partial t}\rho_{\rm{s}}(t)=\frac{1}{(i\hbar)^2}\int_0^t dt^\prime {\rm Tr}_{l,r,u}[H_{\rm{T}}(t),[H_{\rm{T}}(t-t^\prime),\rho_{\rm{tot}}(t^\prime)]],
\end{split}
\end{equation}
where ${\rm Tr}_{l,r,u}$ refers to the trace over each bath degrees of freedom and ${\rm Tr}_{l,r,u}\{ \rho_{\rm{tot}}(t)\}=\rho_{\rm{s}}(t)$ denotes the reduced density operator for the system. We also assume that ${\rm Tr}_{l,r,u}[H_{\rm{T}}(t),\rho_{\rm{tot}}(0)]=0$. Under the Born-Markov approximation, the above equation can be rewritten as~\cite{breuer2002book,gupt2022PRE,shuvadip2022univarsal} 
\begin{equation}\label{A4}
\begin{split}
\dot{\rho_{\rm{s}}}(t)=\frac{1}{(i\hbar)^2}\sum_{\lambda,\lambda^\prime}
\int_0^\infty dt^\prime\ {\rm Tr}_{l,r,u}[H_{\rm{T}}^\lambda(t),[H_{\rm{T}}^{\lambda^{\prime}}(t-t^\prime),\rho_{\rm{s}}(t)\otimes\rho_{l}\otimes\rho_{r}\otimes\rho_{u}]],
\end{split}
\end{equation}
where we use the following properties of the bath operators ${\rm Tr}_\lambda\{{c}_\lambda(t)\rho_\lambda\}=0={\rm Tr}_\lambda\{{c}^\dagger_\lambda(t)\rho_\lambda\}$
and ${\rm Tr}_{l,r,u}\{ [H^\lambda_{\rm{T}}(t),[H^{\lambda^\prime}_{\rm{T}}(t-t'),{\rho_{\rm{s}}}(t)\otimes\rho_{l}\otimes\rho_{r}\otimes\rho_{u}]]\}=0;\lambda\ne\lambda^{\prime};\lambda,\lambda^\prime={l},{r},{u}$. Now, in the above equation, we use the interaction picture system and bath operators
\begin{eqnarray}\label{A5}
    d_{\alpha}(t)&=&e^{{iH_{\rm{S}} t}/{\hbar}}d_{\alpha} e^{{-iH_{\rm{S}} t}/{\hbar}}=\sum_{\omega_\mathbb{ij}>0} e^{{{-i{\omega_\mathbb{ij}}} t}/{\hbar}}d_{\alpha};\quad\quad    {\alpha=\rm{b},\rm{u}}
    \nonumber\\
c_{\lambda}(t)&=&e^{{iH_{\rm{B}} t}/{\hbar}}c_{\lambda} e^{{-iH_{\rm{B}} t}/{\hbar}}=\sum_{k} e^{{{-i({\epsilon_k^{\lambda}-\mu_{\lambda}}) t}/{\hbar}}} c_{\lambda};
\end{eqnarray}
and their hermitian adjoints, where $\omega_\mathbb{ij}$ is defined as the transition energy for the transition between the system eigenstates $|\mathbb{i}\rangle$ and $|\mathbb{j}\rangle$. Eliminating the high-frequency oscillating terms by the standard procedure of secular approximation, one can finally derive the master equation in the following form
\begin{equation}\label{A6}
\dot\rho_{\rm{s}}(t)=\sum_{\lambda}\mathcal{L}_{\lambda}[\rho_{\rm{s}}(t)],
\end{equation}
where the Lindblad operators $\mathcal{L}_{\lambda}[\rho_{\rm{s}}(t)]$ are given by
\begin{eqnarray}\label{A7}
\mathcal{L}_{\lambda}[\rho_{\rm{s}}(t)]=\sum_{\{\omega_{\alpha}\}>0}&\mathcal{G}_{{\lambda}}(\omega_{\alpha})&\left[d_{\alpha}^\dagger(\omega_{\alpha})\rho_{\rm{s}} d_{\alpha}(\omega_{\alpha})-\frac{1}{2}{\{\rho_{\rm{s}}}, d_{\alpha}(\omega_{\alpha})d_{\alpha}^\dagger(\omega_{\alpha})\}\right]
\nonumber\\
+&\mathcal{G}_{{\lambda}}(-\omega_{\alpha})&
\left[d_{\alpha}(\omega_{\alpha})\rho_{\rm{s}} d_{\alpha}^\dagger(\omega_{\alpha})-\frac{1}{2}{\{\rho_{\rm{s}}}, d_{\alpha}^\dagger(\omega_{\alpha})d_{\alpha}(\omega_{\alpha})\}\right].\nonumber\\
\end{eqnarray}
In the above equation, we define the temperature-dependent bath spectral functions as
\begin{equation}\label{A8}
\mathcal{G}_{{\lambda}}(\omega_{\alpha})=\gamma_\lambda(\omega_\alpha)f^+_\lambda(\omega_\alpha);\quad \mathcal{G}_{{\lambda}}(-\omega_{\alpha})=\gamma_\lambda(\omega_\alpha) f^-_\lambda(\omega_\alpha),
\end{equation}
where $\gamma_\lambda(\omega_\alpha)$ is the bare electron transfer rate between the reservoir $\lambda$ and coupled ${\rm QD_\alpha}$. The explicit forms in terms of the system-reservoir coupling constants can be calculated using Fermi's golden rule, as 
$\gamma_\lambda(\omega_\alpha)\equiv\gamma_{\lambda}(\omega_\alpha)=2\pi \sum_{k} |t_{k}^{\alpha\lambda}|^2 \delta\big(\omega_\alpha-\epsilon_{k}^{\lambda}\big)$, where 
$\omega_\alpha$ represents the required amount of energy associated with the transition between ${\rm QD_\alpha}$ and its coupled lead. The function $f^{\pm}_\lambda(\omega_\mathbb{ij})$ represents the Fermi distribution functions (FDF) which are obtained by tracing over the bath density operator, for example, $f_{\lambda}^+\big(\omega_\mathbb{ij}\big)=\Tr_{\lambda}\big(c^\dagger_{\lambda} c_{\lambda} \rho_{\lambda}\big)$, and $f_{\lambda}^-\big(\omega_\mathbb{ij}\big)=\Tr_{\lambda}\big(c_{\lambda} c^\dagger_{\lambda} \rho_{\lambda}\big)$, where the bath operators $c^\dagger_{\lambda}$ and $c_{\lambda}$ obey anti-commutation relation and the reservoir governs the transition between eigenstate $|\mathbb{i}\rangle$ to $|\mathbb{j}\rangle$, that costs $\omega_\mathbb{ij}$ amount of energy. The explicit expressions of the FDFs for various transitions are mentioned in the main text.

Now, to derive the forms of the Lindbladians, we need to express the creation and the annihilation operators in terms of the system eigenstates in the following forms,

\begin{equation}\label{A9}
\begin{split}
\sum_{\{\omega_{\rm{b}}\}}d^\dagger_{\rm{b}}(\omega_{\rm{b}})=d_{\rm{b}}^\dagger(\omega_\mathbb{AB})+d^\dagger_{\rm{b}}(\omega_\mathbb{CD})=|\mathbb{B}\rangle \langle\mathbb{A}|+|\mathbb{D}\rangle \langle\mathbb{C}|\quad&\quad \sum_{\{\omega_{\rm{b}}\}}d_{\rm{b}}(\omega_{\rm{b}})=d_{\rm{b}}(\omega_\mathbb{BA})+d_{\rm{b}}(\omega_\mathbb{DC})=|\mathbb{A}\rangle \langle\mathbb{B}|+|\mathbb{C}\rangle \langle\mathbb{D}|
\\\sum_{\{\omega_{\rm{u}}\}}d_{\rm{u}}^\dagger(\omega_{\rm{u}})=d_{\rm{u}}^\dagger(\omega_\mathbb{AC})+d_{\rm{u}}^\dagger(\omega_\mathbb{BD})=|\mathbb{C}\rangle \langle\mathbb{A}|+|\mathbb{D}\rangle \langle\mathbb{B}|\quad&\quad
\sum_{\{\omega_{\rm{u}}\}}d_{\rm{u}}(\omega_{\rm{u}})=d_{\rm{u}}(\omega_\mathbb{CA})+d_{\rm{u}}(\omega_\mathbb{DB})=|\mathbb{A}\rangle \langle\mathbb{C}|+|\mathbb{B}\rangle \langle\mathbb{D}|
\end{split}
\end{equation}
So, the time evolution of the occupation probabilities which are essentially the diagonal elements of the reduced density matrix $\rho_\mathbb{ii}=\langle\mathbb{i}|\rho_{\rm{s}}(t)|\mathbb{i}\rangle$, can be obtained using the LME [Eq.~\eqref{A7}] as follows,
\begin{equation}\label{A10}
\begin{split}
\dot{\rho}_\mathbb{i}\equiv\frac{d\rho_\mathbb{ii}}{dt}=\langle \mathbb{i}|\dot{\rho_{\rm{s}}}(t)|\mathbb{i}\rangle=\sum_{\lambda}\langle \mathbb{i}|\mathcal{L}_{\lambda}[\rho_{\rm{s}}(t)]|\mathbb{i}\rangle.
    \end{split}
\end{equation}

\section{Explicit expressions of various currents}\label{Appendix-B}

From Eq.~\eqref{S15}, one can evaluate the explicit expressions of both energy and particle currents for each reservoir in the following way
 \begin{equation}\label{EPC}
    \begin{split}
J^l_{\rm{E}}=&\omega_\mathbb{AB}\Gamma_\mathbb{AB}^{l+}+\omega_\mathbb{CD}\Gamma_\mathbb{CD}^{l+}=\varepsilon_{\rm{b}}\Gamma_\mathbb{AB}^{l+}+(\varepsilon_{\rm{b}}+\kappa)\Gamma_\mathbb{CD}^{l+}; \quad 
J_{\rm{E}}^{r}=\omega_\mathbb{AB}\Gamma_\mathbb{AB}^{r+}+\omega_\mathbb{CD}\Gamma_\mathbb{CD}^{r+}=\varepsilon_{\rm{b}}\Gamma_\mathbb{AB}^{r+}+(\varepsilon_{\rm{b}}+\kappa)\Gamma_\mathbb{CD}^{r+};
\\J_{\rm{E}}^{u}=&\omega_\mathbb{AC}\Gamma_\mathbb{AC}^{u+}+\omega_\mathbb{BD}\Gamma_\mathbb{BD}^{u+}=\varepsilon_{\rm{u}}\Gamma_\mathbb{AC}^{u+}+(\varepsilon_{\rm{u}}+\kappa)\Gamma_\mathbb{BD}^{u+}; \quad J_{\rm{N}}^l=\Gamma_\mathbb{AB}^{l+}+\Gamma_\mathbb{CD}^{l+};J_{\rm{N}}^r=\Gamma_\mathbb{AB}^{r+}+\Gamma_\mathbb{CD}^{r+};J_{\rm{N}}^u=\Gamma_\mathbb{AC}^{u+}+\Gamma_\mathbb{BD}^{u+}.
    \end{split}
 \end{equation}
Inserting the above relations, the expression of the heat current associated with each reservoir can be evaluated as
\begin{equation}\label{HC}
\begin{split}
J_{\rm{Q}}^l=&(\varepsilon_{\rm{b}}-\mu_l)\Gamma_\mathbb{AB}^{l+}+(\varepsilon_{\rm{b}}+\kappa-\mu_l)\Gamma_\mathbb{CD}^{l+}; \; 
J_{\rm{Q}}^r=(\varepsilon_{\rm{b}}-\mu_r)\Gamma_\mathbb{AB}^{r+}+(\varepsilon_{\rm{b}}+\kappa-\mu_r)\Gamma_\mathbb{CD}^{r+}; \; 
J_{\rm{Q}}^u=(\varepsilon_{\rm{u}}-\mu_u)\Gamma_\mathbb{AC}^{u+}+(\varepsilon_{\rm{u}}+\kappa-\mu_u)\Gamma_\mathbb{BD}^{u+}.
\end{split}    
\end{equation}

\section{Expression of the steady state transition rate}\label{Appendix-C}
\par To determine the steady-state transition rate, we rewrite Eq.~\eqref{rho2}-\eqref{ME15} as
\begin{equation}\label{matrix}
\mathcal{M}\begin{bmatrix}
\rho_{\mathbb{A}}
\\\rho_{\mathbb{B}}
\\\rho_{\mathbb{C}}
\\\rho_{\mathbb{D}}
\end{bmatrix}=\begin{bmatrix}
0
\\0
\\0
\\1
\end{bmatrix},
\end{equation}
subject to the condition
$\rho_{\mathbb{A}}+\rho_{\mathbb{B}}+\rho_{\mathbb{C}}+\rho_{\mathbb{D}}=1$, and
\begin{equation}\begin{split}
\mathcal{M}=\begin{bmatrix}
-f_u^{1+}-f_{a}^{1+}-f_{r}^{1+} & {f}_{l}^{1-}+{f}_{r}^{1-} & f_u^{1+} & 0
\\f_l^{1+}+f_r^{1+} & -{f}_l^{1-}-{f}_r^{1-}-f_u^{2+} & 0 & {f}_u^{2-}
\\f_u^{1+} & 0 & -f_l^{2+}-f_r^{2+}-{f}_u^{1-} & {f}_l^{2-}+{f}_r^{2-}
\\ 1 & 1 & 1 & 1
\end{bmatrix}
,
\end{split}
\end{equation}
where, for the sake of simplicity of our analysis, we assume that $\gamma_l\simeq\gamma_r\simeq\gamma_u\equiv\gamma$. Solving Eq.~\eqref{matrix} with the above matrix $\mathcal{M}$, one can obtain the steady state population $\{\rho_{\mathbb{i}}\}$ in terms of which we can evaluate the explicit expression
\begin{equation}\label{Gamma}
\Gamma_{\circlearrowright}=-\Gamma_{\circlearrowleft}=\gamma\Bigg[\frac{f_{lr}^{1+}[f_{lr}^{2+}(f_u^{2+}-f_u^{1+})+2f_u^{2+}(f_u^{1+}-1)]-2f_u^{1+}f_{lr}^{2+}(f_u^{2+}-1)}{3f_{lr}^{1+}(f_u^{1+}-f_u^{2+})-6+3f_{lr}^{2+}(f_u^{2+}-f_u^{1+})}\Bigg],
\end{equation}
where, we define $f_{lr}^{1+(2+)}=f_l^{1+(2+)}+f_r^{1+(2+)}$.

\section{Non-negativity of entropy production rate}\label{Appendix-D}

One can evaluate the expression of the entropy production ($\Sigma$) from the entropy change of the system ($\varDelta\mathcal{S}_{\rm{s}}=\mathcal{S}_{\rm{s}}(t)-\mathcal{S}_{\rm{s}}(0)$), which is defined as
\begin{equation}\label{C1}
\begin{split}
\varDelta\mathcal{S}_{\rm{s}}(t)
=-k_B \Tr_{\rm{s}}[\rho_{\rm{s}}(t)\ln{\rho_{\rm{s}}(t)}]+k_B \Tr_{\rm{s}}[\rho_{\rm{s}}(0)\ln{\rho_{\rm{s}}(0)}]
=-k_B \Tr[\rho_{\rm{tot}}(t)\ln{\rho_{\rm{s}}(t)}]+k_B \Tr[\rho_{\rm{tot}}(0)\ln{\rho_{\rm{s}}(0)}].
\end{split}
\end{equation}
We assume that the initial equilibrium state, $\rho_{\rm{tot}}(0)$, does not display any entanglement or correlation between the system and the environment. Therefore
\begin{equation}\label{C2}
\rho_{\rm{tot}}(0)=\rho_{\rm{s}}(0)\prod_{\lambda}\rho_{\rm{\lambda}}^{\rm{eq}}.
\end{equation}
Inserting Eq.~\eqref{C2} into Eq.~\eqref{C1}, we can continue as follows
\begin{equation}\label{C3}
\begin{split}
 \varDelta\mathcal{S}_{\rm{s}}(t)=&-k_B \Tr[\rho_{\rm{tot}}(t)\ln{\rho_{\rm{s}}(t)}]+k_B \Tr[\rho_{\rm{tot}}(0)\ln{\rho_{\rm{tot}}(0)}]-k_B \sum_{\lambda}\Tr[\rho_{\rm{tot}}(0)\ln{\rho_{\lambda}^{\rm{eq}}}]\\
 =&-k_B \Tr[\rho_{\rm{tot}}(t)\ln\left\{{\rho_{\rm{s}}(t)}\prod_{\lambda}\rho_{\lambda}^{\rm{eq}}\right\}]+k_B \Tr[\rho_{\rm{tot}}(0)\ln{\rho_{\rm{tot}}(0)}]+k_B \sum_{\lambda}\Tr_{\lambda}[\{\rho_{\lambda}(t)-\rho_{\lambda}^{\rm{eq}}\}\ln{\rho_{\lambda}^{\rm{eq}}}].
\end{split}
\end{equation}
Again, $\rho_{\rm{tot}}(t)$ and $\rho_{\rm{tot}}(0)$ are related through unitary evolution $\rho_{\rm{tot}}(t)=\mathcal{U}\rho_{\rm{tot}}(0)\mathcal{U}^\dagger$, which implies $\Tr[\rho_{\rm{tot}}(t)\ln\rho_{\rm{tot}}(t)]=\Tr[\rho_{\rm{tot}}(0)\ln{\rho_{\rm{tot}}(0)}]$. 
Applying this relation in the above equation, the final expression of the entropy change of the system can be calculated as
\begin{equation}\label{C5}
    \varDelta\mathcal{S}_{\rm{s}}(t)=-k_B \Tr[\rho_{\rm{tot}}(t)\ln\left\{{\rho_{\rm{s}}(t)}\prod_{\lambda}\rho_{\lambda}^{\rm{eq}}\right\}]+k_B \Tr[\rho_{\rm{tot}}(t)\ln{\rho_{\rm{tot}}(t)}]+k_B \sum_{\lambda}\Tr_{\lambda}[\{\rho_{\lambda}(t)-\rho_{\lambda}^{\rm{eq}}\}\ln{\rho_{\lambda}^{\rm{eq}}}].
\end{equation}
The last term of the above equation can be identified as the \textit{entropy flow} ($\Phi$), representing the reversible contribution to the system entropy change due to heat exchange with the reservoirs. A comparison of the above equation with Eq.~\eqref{EP2} defines the \textit{entropy production}, representing the irreversible contribution to the entropy change of the system and the \textit{entropy flow} as
\begin{equation}\label{C6}
\Sigma(t)=k_B\Tr[\rho_{\rm{tot}}(t)\ln\{\rho_{\rm{tot}}(t)\}]-k_B\Tr[\rho_{\rm{tot}}(t)\ln\left\{\rho_{\rm{s}}(t)\prod_{\lambda}\rho_{\lambda}^{\rm{eq}}\right\}]; \qquad
\Phi(t)=k_B\sum_{\lambda}\Tr_{\rm{\lambda}}\left[\left\{\rho_{\lambda}(t)-\rho_{\lambda}^{\rm{eq}}\right\}\ln{\rho_{\lambda}^{\rm{eq}}}\right].
\end{equation}
The above expression of the \textit{entropy production} can be expressed in terms of the relative entropy
\begin{equation}\label{C7}
\Sigma(t)\equiv\mathcal{D}\Bigg[\rho_{\rm{tot}}(t)||\left\{\rho_{\rm{s}}(t)\prod_{\lambda}\rho_{\lambda}^{\rm{eq}}\right\}\Bigg],
\end{equation}
where $\mathcal{D}[\rho||{\rho^\prime}]$ is the quantum relative entropy between two density matrices $\rho$ and ${\rho^\prime}$, defined via
\begin{equation}\label{C8}
\mathcal{D}[\rho||{\rho^\prime}]:=\Tr[\rho\ln\rho]-\Tr[\rho\ln{\rho^\prime}].    
\end{equation}
The non-negativity of relative entropy is affirmed, attaining a value of zero solely in the case of complete matrix identity. While the non-negativity of entropy production doesn't imply the same for its rate, in the limit of large reservoirs, $\Sigma(t)$ is expected to converge to a convex, monotonically increasing function of time \cite{esposito2010entropy}. In the same limit, if the system dynamics are described by a Markovian quantum Lindblad master equation, implying entropy production as a convex functional of the system density matrix \cite{breuer2002book,esposito2010entropy}, the entropy production rate $\dot{\Sigma}(t)$ would eventually be positive, only reaching zero for the equilibrium state.

\section{Derivation of the microscopic definition of the entropy production rate}\label{Appendix-E}

Considering the von Neumann entropy defined as $\mathcal{S}_{\rm{s}}(t)=-k_B \sum_{\mathbb{i}}\rho_{\mathbb{i}}(t)\ln{\rho_{\mathbb{i}}(t)}$, where, $\rho_{\mathbb{i}}$ signifies populations of the system eigenstates ($\mathbb{i=A,B,C,D}$), the change in the system entropy is given by
\begin{equation}
\begin{split}
\varDelta\mathcal{S}_{\rm{s}}(t)=&\mathcal{S}_{\rm{s}}(t)-\mathcal{S}_{\rm{s}}(0)
 =-k_B \sum_{\mathbb{i}}\rho_{\mathbb{i}}(t)\ln{\rho_{\mathbb{i}}(t)}+k_B \sum_{\mathbb{i}}\rho_{\mathbb{i}}(0)\ln{\rho_{\mathbb{i}}(0)}.
 \end{split}
\end{equation}
So, the time evolution of the entropy change can be evaluated as
\begin{equation}
\begin{split}
\frac{d}{dt}\varDelta\mathcal{S}_{\rm{s}}(t)=&-k_B \sum_{\mathbb{i}}\dot{\rho_{\mathbb{i}}}(t)\ln{\rho_{\mathbb{i}}(t)}+k_B \sum_{\mathbb{i}}\dot{\rho_{\mathbb{i}}}(0)\ln{\rho_{\mathbb{i}}(0)}
=-k_B \sum_{\mathbb{i}}\dot{\rho_{\mathbb{i}}}(t)\ln{\rho_{\mathbb{i}}(t)}
\equiv-k_B \sum_{\mathbb{i}}\dot{\rho_{\mathbb{i}}}\ln{\rho_{\mathbb{i}}}.
 \end{split}
\end{equation}
Now, using Eq.~\eqref{rho2} for $\dot{\rho_{\mathbb{i}}}$, we recover Eq.~\eqref{delta-s-dot} of the main text:
\begin{equation}
\begin{split}
\frac{d}{dt}\varDelta\mathcal{S}_{\rm{s}}(t)
=k_B\Bigg[\Gamma^{l+}_\mathbb{AB}\ln(\frac{\rho_{\mathbb{A}}}{\rho_{\mathbb{B}}})+\Gamma^{r+}_\mathbb{AB}\ln(\frac{\rho_{\mathbb{A}}}{\rho_{\mathbb{B}}})+\Gamma^{u+}_\mathbb{BD}\ln(\frac{\rho_{\mathbb{B}}}{\rho_{\mathbb{D}}})
+\Gamma^{l-}_\mathbb{DC}\ln(\frac{\rho_{\mathbb{D}}}{\rho_{\mathbb{C}}})+\Gamma^{r-}_\mathbb{DC}\ln(\frac{\rho_{\mathbb{D}}}{\rho_{\mathbb{C}}})+\Gamma^{u-}_\mathbb{CA}\ln(\frac{\rho_{\mathbb{C}}}{\rho_{\mathbb{A}}})\Bigg].  
\end{split}
\end{equation}
By comparing the above equation with Eq.~\eqref{EP2}, we obtain Eq.~\eqref{sigma-phi-gen} for the general expressions of $\dot{\Sigma}(t)$ and $\dot{\Phi}(t)$:
\begin{equation}
\begin{split}
\dot{\Sigma}(t)=&k_B\Bigg[\Gamma^{l+}_\mathbb{AB}\ln(\frac{{\rm{k}}_\mathbb{AB}^{l+}\rho_{\mathbb{A}}}{{\rm{k}}_\mathbb{BA}^{l-}\rho_{\mathbb{B}}})+\Gamma^{r+}_\mathbb{AB}\ln(\frac{{\rm{k}}_\mathbb{AB}^{r+}\rho_{\mathbb{A}}}{{\rm{k}}_\mathbb{BA}^{r-}\rho_{\mathbb{B}}})+\Gamma^{u+}_\mathbb{BD}\ln(\frac{{\rm{k}}_\mathbb{BD}^{u+}\rho_{\mathbb{B}}}{{\rm{k}}_\mathbb{DB}^{u-}\rho_{\mathbb{D}}})
+\Gamma^{l-}_\mathbb{DC}\ln(\frac{{\rm{k}}_\mathbb{DC}^{l-}\rho_{\mathbb{D}}}{{\rm{k}}_\mathbb{CD}^{l+}\rho_{\mathbb{C}}})+\Gamma^{r-}_\mathbb{DC}\ln(\frac{{\rm{k}}_\mathbb{DC}^{r-}\rho_{\mathbb{D}}}{{\rm{k}}_\mathbb{CD}^{r+}\rho_{\mathbb{C}}})+\Gamma^{u-}_\mathbb{CA}\ln(\frac{{\rm{k}}_\mathbb{CA}^{u-}\rho_{\mathbb{C}}}{{\rm{k}}_\mathbb{AC}^{u+}\rho_{\mathbb{A}}})\Bigg]\\
=&k_B\Bigg[({\rm{k}}_\mathbb{AB}^{l+}\rho_{\mathbb{A}}-{\rm{k}}_\mathbb{BA}^{l-}\rho_{\mathbb{B}})\ln(\frac{{\rm{k}}_\mathbb{AB}^{l+}\rho_{\mathbb{A}}}{{\rm{k}}_\mathbb{BA}^{l-}\rho_{\mathbb{B}}})+({\rm{k}}_\mathbb{AB}^{r+}\rho_{\mathbb{A}}-{\rm{k}}_\mathbb{BA}^{r-}\rho_{\mathbb{B}})\ln(\frac{{\rm{k}}_\mathbb{AB}^{r+}\rho_{\mathbb{A}}}{{\rm{k}}_\mathbb{BA}^{r-}\rho_{\mathbb{B}}})+({\rm{k}}_\mathbb{BD}^{u+}\rho_{\mathbb{B}}-{\rm{k}}_\mathbb{DB}^{u-}\rho_{\mathbb{D}})\ln(\frac{{\rm{k}}_\mathbb{BD}^{u+}\rho_{\mathbb{B}}}{{\rm{k}}_\mathbb{DB}^{u-}\rho_{\mathbb{D}}})\\
+&({\rm{k}}_\mathbb{DC}^{l-}\rho_{\mathbb{D}}-{\rm{k}}_\mathbb{CD}^{l+}\rho_{\mathbb{C}})\ln(\frac{{\rm{k}}_\mathbb{DC}^{l-}\rho_{\mathbb{D}}}{{\rm{k}}_\mathbb{CD}^{l+}\rho_{\mathbb{C}}})+({\rm{k}}_\mathbb{DC}^{r-}\rho_{\mathbb{D}}-{\rm{k}}_\mathbb{CD}^{r+}\rho_{\mathbb{C}})\ln(\frac{{\rm{k}}_\mathbb{DC}^{r-}\rho_{\mathbb{D}}}{{\rm{k}}_\mathbb{CD}^{r+}\rho_{\mathbb{C}}})+({\rm{k}}_\mathbb{CA}^{u-}\rho_{\mathbb{C}}-{\rm{k}}_\mathbb{AC}^{u+}\rho_{\mathbb{A}})\ln(\frac{{\rm{k}}_\mathbb{CA}^{u-}\rho_{\mathbb{C}}}{{\rm{k}}_\mathbb{AC}^{u+}\rho_{\mathbb{A}}})\Bigg],\\
\dot{\Phi}(t)=&-k_B\Bigg[\Gamma^{l+}_\mathbb{AB}\ln(\frac{{\rm{k}}_\mathbb{AB}^{l+}}{{\rm{k}}_\mathbb{BA}^{l-}})+\Gamma^{r+}_\mathbb{AB}\ln(\frac{{\rm{k}}_\mathbb{AB}^{r+}}{{\rm{k}}_\mathbb{BA}^{r-}})+\Gamma^{u+}_\mathbb{BD}\ln(\frac{{\rm{k}}_\mathbb{BD}^{u+}}{{\rm{k}}_\mathbb{DB}^{u-}})
+\Gamma^{l-}_\mathbb{DC}\ln(\frac{{\rm{k}}_\mathbb{DC}^{l-}}{{\rm{k}}_\mathbb{CD}^{l+}})+\Gamma^{r-}_\mathbb{DC}\ln(\frac{{\rm{k}}_\mathbb{DC}^{r-}}{{\rm{k}}_\mathbb{CD}^{r+}})+\Gamma^{u-}_\mathbb{CA}\ln(\frac{{\rm{k}}_\mathbb{CA}^{u-}}{{\rm{k}}_\mathbb{AC}^{u+}})\Bigg].
\end{split}
\end{equation}
There is no net entropy change in the system at the steady state, which reduces Eq.~\eqref{ss-sigma} for the form of the entropy production rate:
\begin{equation}
\begin{split}
\dot{\Sigma}(t)&=-\dot{\Phi}(t)
=k_B\Bigg[\Gamma^{l+}_\mathbb{AB}\ln(\frac{{{\rm{k}}}_\mathbb{AB}^{l+}}{{\rm{k}}_\mathbb{BA}^{l-}})+\Gamma^{r+}_\mathbb{AB}\ln(\frac{{\rm{k}}_\mathbb{AB}^{r+}}{{\rm{k}}_\mathbb{BA}^{r-}})+\Gamma^{u+}_\mathbb{BD}\ln(\frac{{\rm{k}}_\mathbb{BD}^{u+}}{{\rm{k}}_\mathbb{DB}^{u-}})
+\Gamma^{l-}_\mathbb{DC}\ln(\frac{{\rm{k}}_\mathbb{DC}^{l-}}{{\rm{k}}_\mathbb{CD}^{l+}})+\Gamma^{r-}_\mathbb{DC}\ln(\frac{{\rm{k}}_\mathbb{DC}^{r-}}{{\rm{k}}_\mathbb{CD}^{r+}})+\Gamma^{u-}_\mathbb{CA}\ln(\frac{{\rm{k}}_\mathbb{CA}^{u-}}{{\rm{k}}_\mathbb{AC}^{u+}})\Bigg]\\
=&k_B\Bigg[\Gamma^{lr+}_\mathbb{AB}\ln(\frac{{\rm{k}}_\mathbb{AB}^{l+}}{{\rm{k}}_\mathbb{BA}^{l-}})+\Gamma^{u+}_\mathbb{BD}\ln(\frac{{\rm{k}}_\mathbb{BD}^{u+}}{{\rm{k}}_\mathbb{DB}^{u-}})
+\Gamma^{lr-}_\mathbb{DC}\ln(\frac{{\rm{k}}_\mathbb{DC}^{l-}}{{\rm{k}}_\mathbb{CD}^{l+}})+\Gamma^{u-}_\mathbb{CA}\ln(\frac{{\rm{k}}_\mathbb{CA}^{u-}}{{\rm{k}}_\mathbb{AC}^{u+}})\Bigg]+k_B\Bigg[\Gamma^{r+}_\mathbb{AB}\ln(\frac{{\rm{k}}_\mathbb{AB}^{r+}{\rm{k}}_\mathbb{BA}^{l-}}{{\rm{k}}_\mathbb{BA}^{r-}{\rm{k}}_\mathbb{AB}^{l+}})+\Gamma^{r-}_\mathbb{DC}\ln(\frac{{\rm{k}}_\mathbb{DC}^{r-}{\rm{k}}_\mathbb{CD}^{l+}}{{\rm{k}}_\mathbb{CD}^{r+}{\rm{k}}_\mathbb{DC}^{l-}})\Bigg]\\
=&k_B \Gamma_{\circlearrowright} \ln(\frac{{\rm{k}}_\mathbb{AB}^{l+}{\rm{k}}_\mathbb{BD}^{u+}{\rm{k}}_\mathbb{DC}^{l-}{\rm{k}}_\mathbb{CA}^{u-}}{{\rm{k}}_\mathbb{BA}^{l-}{\rm{k}}_\mathbb{DB}^{u-}{\rm{k}}_\mathbb{CD}^{l+}{\rm{k}}_\mathbb{AC}^{u+}})+k_B (\Gamma^{r+}_\mathbb{AB}-\Gamma^{r-}_\mathbb{DC})\ln(\frac{{\rm{k}}_\mathbb{AB}^{r+}{\rm{k}}_\mathbb{BA}^{l-}}{{\rm{k}}_\mathbb{BA}^{r-}{\rm{k}}_\mathbb{AB}^{l+}})+k_B \Gamma^{r-}_\mathbb{DC} \ln(\frac{{\rm{k}}_\mathbb{DC}^{r-}{\rm{k}}_\mathbb{CD}^{l+}{\rm{k}}_\mathbb{AB}^{r+}{\rm{k}}_\mathbb{BA}^{l-}}{{\rm{k}}_\mathbb{BA}^{r-}{\rm{k}}_\mathbb{AB}^{l+}{\rm{k}}_\mathbb{CD}^{r+}{\rm{k}}_\mathbb{DC}^{l-}})\\
=&\kappa\Gamma_{\circlearrowright} \Bigg[\left(\frac{k_B}{\kappa}\right)\ln(\frac{{\rm{k}}_\mathbb{AB}^{l+}{\rm{k}}_\mathbb{BD}^{u+}{\rm{k}}_\mathbb{DC}^{l-}{\rm{k}}_\mathbb{CA}^{u-}}{{\rm{k}}_\mathbb{BA}^{l-}{\rm{k}}_\mathbb{DB}^{u-}{\rm{k}}_\mathbb{CD}^{l+}{\rm{k}}_\mathbb{AC}^{u+}})\Bigg]
+\{\varepsilon_{\rm{b}}\Gamma^{r+}_\mathbb{AB}-(\varepsilon_{\rm{b}}+\kappa)\Gamma^{r-}_\mathbb{DC}\}\Bigg[\left(\frac{k_B}{\kappa}\right)\ln(\frac{{\rm{k}}_\mathbb{BA}^{r-}{\rm{k}}_\mathbb{AB}^{l+}{\rm{k}}_\mathbb{CD}^{r+}{\rm{k}}_\mathbb{DC}^{l-}}{{\rm{k}}_\mathbb{DC}^{r-}{\rm{k}}_\mathbb{CD}^{l+}{\rm{k}}_\mathbb{AB}^{r+}{\rm{k}}_\mathbb{BA}^{l-}})\Bigg]\\
+&(\Gamma^{r+}_\mathbb{AB}-\Gamma^{r-}_\mathbb{DC})\Bigg[k_B (1+\theta)\ln(\frac{{\rm{k}}_\mathbb{AB}^{r+}{\rm{k}}_\mathbb{BA}^{l-}}{{\rm{k}}_\mathbb{BA}^{r-}{\rm{k}}_\mathbb{AB}^{l+}})+k_B \theta\ln(\frac{{\rm{k}}_\mathbb{DC}^{r-}{\rm{k}}_\mathbb{CD}^{l+}}{{\rm{k}}_\mathbb{CD}^{r+}{\rm{k}}_\mathbb{DC}^{l-}})\Bigg].
\end{split}
\end{equation}

\section{Determination of the sign of $\Gamma_{\circlearrowright}$ from $\mathcal{P}\mathcal{Q}^{-1}$ and criteria of Pseudo-ICC}\label{Appendix-F}

\subsection*{\textbf{Expression of $\mathcal{P}\mathcal{Q}^{-1}$}}

From Eq.~\eqref{force-e-r}, the condition $\mathcal{F}_{\rm{E}}^u=0$ implies   
\begin{equation}\label{R1}  \left(\frac{{\rm{k}}_\mathbb{AB}^{a+}{\rm{k}}_\mathbb{BD}^{r+}{\rm{k}}_\mathbb{DC}^{a-}{\rm{k}}_\mathbb{CA}^{r-}}{{\rm{k}}_\mathbb{BA}^{a-}{\rm{k}}_\mathbb{DB}^{r-}{\rm{k}}_\mathbb{CD}^{a+}{\rm{k}}_\mathbb{AC}^{r+}}\right)=1.
\end{equation}
We can rewrite the above condition as
\begin{equation}\label{R2}
\left(\frac{\rm{k}_\mathbb{AB}^{ab+}\rm{k}_\mathbb{BD}^{r+}\rm{k}_\mathbb{DC}^{ab-}\rm{k}_\mathbb{CA}^{r-}}{\rm{k}_\mathbb{BA}^{ab-}\rm{k}_\mathbb{DB}^{r-}\rm{k}_\mathbb{CD}^{ab+}\rm{k}_\mathbb{AC}^{r+}}\right)= \left(\frac{\rm{k}_\mathbb{BA}^{a+}\rm{k}_\mathbb{AB}^{ab+}\rm{k}_\mathbb{CD}^{a+}\rm{k}_\mathbb{DC}^{ab-}}{\rm{k}_\mathbb{AB}^{a-}\rm{k}_\mathbb{BA}^{ab-}\rm{k}_\mathbb{DC}^{a-}\rm{k}_\mathbb{CD}^{ab+}}\right).
\end{equation}
The r.h.s. of the above equation can be expressed as
\begin{equation}\label{R3}
\begin{split}
\text{r.h.s.}=&\left(\frac{\rm{k}_\mathbb{BA}^{a+}\rm{k}_\mathbb{AB}^{ab+}\rm{k}_\mathbb{CD}^{a+}\rm{k}_\mathbb{DC}^{ab-}}{\rm{k}_\mathbb{AB}^{a-}\rm{k}_\mathbb{BA}^{ab-}\rm{k}_\mathbb{DC}^{a-}\rm{k}_\mathbb{CD}^{ab+}}\right)\\
=&\left\{\frac{1+({\rm{k}_\mathbb{AB}^{b+}}/{\rm{k}_\mathbb{AB}^{a+}})}{1+({\rm{k}_\mathbb{BA}^{b-}/\rm{k}_\mathbb{BA}^{a-}})}\right\}\left\{\frac{1+({\rm{k}_\mathbb{CD}^{b+}}/{\rm{k}_\mathbb{CD}^{a+}})}{1+({\rm{k}_\mathbb{DC}^{b-}/\rm{k}_\mathbb{DC}^{a-}})}\right\}^{-1}\equiv\mathcal{P}\mathcal{Q}^{-1}.
\end{split}
\end{equation}
One can determine the sign of $\Gamma_{\circlearrowright}$ from $\mathcal{P}\mathcal{Q}^{-1}$.

\subsection*{\textbf{Determination of the sign of $\Gamma_{\circlearrowright}$ from $\mathcal{P}\mathcal{Q}^{-1}$}}

When, $\mathcal{P}=\mathcal{Q}$, applying Eq.~\eqref{R3} to Eq.~\eqref{R2}, we find the l.h.s of Eq.~\eqref{R2} as:
\begin{equation}\label{R4}
\begin{split}
\text{l.h.s.}=&\left(\frac{\rm{k}_\mathbb{AB}^{ab+}\rm{k}_\mathbb{BD}^{r+}\rm{k}_\mathbb{DC}^{ab-}\rm{k}_\mathbb{CA}^{r-}}{\rm{k}_\mathbb{BA}^{ab-}\rm{k}_\mathbb{DB}^{r-}\rm{k}_\mathbb{CD}^{ab+}\rm{k}_\mathbb{AC}^{r+}}\right)=1,
\end{split}
\end{equation}
which can be rewritten as
\begin{equation}\label{R5}
\left(\frac{\rm{k}_\mathbb{AB}^{ab+}\rho_{\mathbb{A}}}{\rm{k}_\mathbb{BA}^{ab-}\rho_{\mathbb{B}}}\right)\left(\frac{\rm{k}_\mathbb{BD}^{r+}\rho_{\mathbb{B}}}{\rm{k}_\mathbb{DB}^{r-}\rho_{\mathbb{D}}}\right)\left(\frac{\rm{k}_\mathbb{DC}^{ab-}\rho_{\mathbb{D}}}{\rm{k}_\mathbb{CD}^{ab+}\rho_{\mathbb{C}}}\right)\left(\frac{\rm{k}_\mathbb{CA}^{r-}\rho_{\mathbb{C}}}{\rm{k}_\mathbb{AC}^{r+}\rho_{\mathbb{A}}}\right)=1.   
\end{equation}
To satisfy this condition at the steady-state, following Eq.~\eqref{gamma}, each term within the parentheses must equal $1$. When combined with Eqs.~\eqref{ME14}, this in turn leads to $\Gamma_{\circlearrowright}=\Gamma_{\circlearrowleft}=0$. 

On the other hand, following the same procedure, we can verify that for $\mathcal{P}\mathcal{Q}^{-1}\gtrless1$, we have
\begin{equation}\label{R6}
\left(\frac{\rm{k}_\mathbb{AB}^{ab+}\rho_{\mathbb{A}}}{\rm{k}_\mathbb{BA}^{ab-}\rho_{\mathbb{B}}}\right)\left(\frac{\rm{k}_\mathbb{BD}^{r+}\rho_{\mathbb{B}}}{\rm{k}_\mathbb{DB}^{r-}\rho_{\mathbb{D}}}\right)\left(\frac{\rm{k}_\mathbb{DC}^{ab-}\rho_{\mathbb{D}}}{\rm{k}_\mathbb{CD}^{ab+}\rho_{\mathbb{C}}}\right)\left(\frac{\rm{k}_\mathbb{CA}^{r-}\rho_{\mathbb{C}}}{\rm{k}_\mathbb{AC}^{r+}\rho_{\mathbb{A}}}\right)\gtrless1.  
\end{equation}
At steady-state, when each term in parentheses is the same, the application of Eqs.~\eqref{ME14}~and~\eqref{gamma} yields the result $\Gamma_{\circlearrowright}\lessgtr0$.

\subsubsection*{\textbf{Consistency check: zero force - zero current}}

As a simple consistency check, let us consider the trivial case where both forces vanish, $(\mathcal{F}_{\rm{E}}^{r},\mathcal{F}_{\rm{N}}^{r})=0$. In this limit, both currents $(J_{\rm{E}}^r, J_{\rm{N}}^r)$ must be zero.  

This is evident since $\mathcal{F}_{\rm E}^r = 0$, Eq.~\eqref{force-e-b} gives 
$\mathcal{M} = \mathcal{N}$. Likewise, $\mathcal{F}_{\rm N}^r = 0$ implies 
    $\mathcal{M} = 1$ from Eq.~\eqref{force-n-b}. Thus, the condition 
    $(\mathcal{F}_{\rm E}^r, \mathcal{F}_{\rm N}^r) = 0$ is equivalent to 
    $\mathcal{M} = \mathcal{N} = 1$, and hence $\mathcal{X} = \mathcal{Y} = 0$. Similarly, for $\mathcal{F}_{\rm E}^r = 0$, the Fermi functions further imply 
    $\mathcal{P}\mathcal{Q}^{-1} = 1$, consistent with $\mathcal{M} = 1$. 
    This yields $\Gamma_{\circlearrowright} = 0$. 

Therefore, $\mathcal{X} = \mathcal{Y} = \Gamma_{\circlearrowright} = 0$ and Eq.~\eqref{R8} gives $\Gamma_{\mathbb{AB}}^{r+} = \Gamma_{\mathbb{CD}}^{r+} = 0$. So, both currents $(J_{\rm E}^r, J_{\rm N}^r)$ vanish, as expected.

\subsubsection*{\textbf{Setting $\mathcal{F}_{\rm{E}}^{r}=0$: Pseudo-ICC in $J_{\rm{E}}^r$}}

In this case, $\mathcal{F}_{\rm{E}}^{r}=0$ and $\mathcal{F}_{\rm{N}}^{r}>0$. From Eq.~\eqref{force-e-b}, the condition $\mathcal{F}_{\rm{E}}^{r}=0$ implies $\mathcal{M}=\mathcal{N}$, while Eq.~\eqref{force-n-b} shows that $\mathcal{F}_{\rm{N}}^{r}>0$ requires $\mathcal{M}>1$. Thus, we obtain $\mathcal{M}=\mathcal{N}>1$, which means $\mathcal{X}>0$ and $\mathcal{Y}>0$, and consequently $J_{\rm N}^r = \frac{1}{2}(\mathcal{X} + \mathcal{Y})> 0$. This agrees with the non-negativity of the entropy production rate, $\dot{\Sigma} = J_{\rm N}^r \mathcal{F}_{\rm N}^r$. Hence, the spin-polarized particle current remains positive [dotted red line in Fig.~\ref{RM2}(a)].

Since, the particle current is positive in this case, the energy current satisfies $J_{\rm{E}}^r>\kappa\Gamma_\mathbb{CD}^{r+}$ [Cf.~Eq.~\eqref{flux}]. With $\mathcal{F}_{\rm{E}}^{r}=0$ and $\mathcal{M}>1$, it follows that $\Gamma_{\circlearrowright}\lessgtr0$ for $\kappa\gtrless0$, which in turn implies [Cf. Eq.~\eqref{R8}] that $\Gamma_\mathbb{CD}^{r+}>0$ for $\kappa>0$, and $\Gamma_\mathbb{CD}^{r+}\gtrless0$ for $\kappa<0$. Thus, for $\kappa>0$ , we immediately obtain $J_{\rm{E}}^r>0$, while for $\kappa<0$ we must examine two sub-cases: (a)~$\varepsilon_{\rm{b}}+\kappa>0$ and (b)~$\varepsilon_{\rm{b}}+\kappa <0$.
    
    \begin{itemize}
        \item For case (a), it is straightforward to show that, $J_{\rm{E}}^r=\varepsilon_{\rm{b}}\Gamma_\mathbb{AB}^{r+}+(\varepsilon_{\rm{b}}+\kappa)\Gamma_\mathbb{CD}^{r+}>0$, when $\Gamma_\mathbb{CD}^{r+}>0$ and that $J_{\rm{E}}^r>\kappa\Gamma_\mathbb{CD}^{r+}>0$, when $\Gamma_\mathbb{CD}^{r+}<0$.

        \item In contrast, for case (b), $J_{\rm{E}}^r$ can be either positive or negative [Cf.~Eq.~\eqref{flux}], indicating the possibility of pseudo-ICC in $J_{\rm{E}}^r$ [dashed blue-green line FIG.~\ref{RM2}(a)].
    \end{itemize}

\subsubsection*{\textbf{Setting $\mathcal{F}_{\rm{N}}^{r}=0$: Pseudo-ICC in $J_{\rm{N}}^r$}}

In this case, $\mathcal{F}_{\rm N}^r = 0$ and $\mathcal{F}_{\rm E}^r > 0$. 
    From Eq.~\eqref{force-n-b}, these conditions immediately give $\mathcal{M} > 1$. 
    Moreover, using $\mathcal{F}_{\rm N}^r = 0$ in Eq.~\eqref{force-n-b-2}, we obtain
    \begin{equation}\label{R9}
        (\varepsilon_{\rm b} + \kappa)\ln \mathcal{M} 
        = \varepsilon_{\rm b} \ln \mathcal{N}.
    \end{equation}
Substituting $\mathcal{M} > 1$ into this relation leads to different outcomes depending on the sign of $\kappa$, as summarized below:
\begin{itemize}
    \item For $\kappa > 0$ and case (a) with $\kappa < 0$, but $\varepsilon_{\rm b} + \kappa > 0$, then,  
    Eq.~\eqref{R9} gives $\mathcal{N} > 1$. Thus, the microscopic condition becomes 
    $\mathcal{M} > 1$ and $\mathcal{N} > 1$, which implies $\mathcal{X} > 0$ and 
    $\mathcal{Y} > 0$ [cf.~Eq.~\eqref{R7}], leading to $J_{\rm N}^r = \tfrac{1}{2}(\mathcal{X} + \mathcal{Y}) > 0$.

    \item In contrast, for case (b) (i.e., $\varepsilon_{\rm b} + \kappa < 0$),  
    Eq.~\eqref{R9} yields $\mathcal{N} < 1$. Here, the eigenstates 
    $|\mathbb{C}\rangle$ and $|\mathbb{D}\rangle$ flip together, giving 
    $\mathcal{M} > 1$ and $\mathcal{N} < 1$, and hence 
    $\mathcal{X} > 0$ and $\mathcal{Y} < 0$. Consequently, $
        J_{\rm N}^r = \tfrac{1}{2}(\mathcal{X} + \mathcal{Y}) \gtrless 0$, meaning the spin-polarized particle current can be positive or negative,
    depending on the balance between the positive $\mathcal{X}$ and negative 
    $\mathcal{Y}$. This allows for pseudo-ICC in $J_{\rm{N}}^r$ [dashed red line in 
    Fig.~\ref{RM2}(b)].
\end{itemize}

Again, since $\mathcal{F}_{\rm E}^{r}>0$ and $\mathcal{M}>1$, it follows that 
$\Gamma_{\circlearrowright}\lessgtr0$ for $\kappa\gtrless0$, leading to the following case-wise outcomes:

\begin{itemize}

     \item \textbf{For $\kappa>0$:}  
    The particle current is positive, allowing us to rewrite the energy current as 
    $J_{\rm E}^r>\kappa\Gamma_{\mathbb{CD}}^{r+}$.  
    Because $\mathcal{X}>0$ and $\mathcal{Y}>0$, Eq.~\eqref{R8} gives 
    $\Gamma_{\mathbb{CD}}^{r+}>0$ (as $\Gamma_{\circlearrowright}<0$). Hence, $J_{\rm E}^r>\kappa\Gamma_{\mathbb{CD}}^{r+}>0$.

     \item \textbf{Case (a): $\varepsilon_{\rm b}+\kappa>0$ with $\kappa<0$:}  
    Here $\Gamma_{\mathbb{CD}}^{r+}\gtrless0$ (because $\Gamma_{\circlearrowright}>0$ and $\mathcal{Y}>0$). Consequently, $J_{\rm E}^r=\varepsilon_{\rm b}\Gamma_{\mathbb{AB}}^{r+}
        +(\varepsilon_{\rm b}+\kappa)\Gamma_{\mathbb{CD}}^{r+}>0$,
    whenever $\Gamma_{\mathbb{CD}}^{r+}>0$, and $J_{\rm E}^r>\kappa\Gamma_{\mathbb{CD}}^{r+}>0$
    when $\Gamma_{\mathbb{CD}}^{r+}<0$.

     \item \textbf{Case (b): $\varepsilon_{\rm b}+\kappa<0$:} 
    With $\Gamma_{\circlearrowright}>0$, $\mathcal{X}>0$, and $\mathcal{Y}<0$, Eq.~\eqref{R8} yields $\Gamma_{\mathbb{AB}}^{r+}>0, \quad \Gamma_{\mathbb{CD}}^{r+}<0$. Therefore, $J_{\rm E}^r=\varepsilon_{\rm b}\Gamma_{\mathbb{AB}}^{r+}
        +(\varepsilon_{\rm b}+\kappa)\Gamma_{\mathbb{CD}}^{r+}>0$.
    \end{itemize}

\end{document}